\documentclass[acmsmall]{acmart}

\AtBeginDocument{%
  \providecommand\BibTeX{{%
    \normalfont B\kern-0.5em{\scshape i\kern-0.25em b}\kern-0.8em\TeX}}}

\renewcommand{\footnotesize}{\fontsize{5}{6}\selectfont}

\usepackage{graphicx}
\usepackage{float}
\usepackage{amsfonts}
\usepackage{indentfirst}
\usepackage{enumitem}
\usepackage{amsmath}
\usepackage{color}
\usepackage{subcaption}
\definecolor{darkgreen}{RGB}{47,109,79}
\definecolor{darkblue}{RGB}{57,79,99}
\usepackage{hyperref}
\hypersetup{bookmarks=false, colorlinks=true, plainpages=false, linkcolor=darkgreen, citecolor=darkgreen, urlcolor=darkblue, filecolor=blue}

\setcopyright{none}
\renewcommand\footnotetextcopyrightpermission[1]{} 
\usepackage[marginal,hang]{footmisc}
\renewcommand{\footnoterule}{%
  \kern -3pt
  \hrule width 1in 
  \kern 2pt
}

\makeatletter
\def\url@leostyle{%
  \@ifundefined{selectfont}{\def\UrlFont{}}%
  {\def\UrlFont{}}%
}
\makeatother  
\makeatletter
\def\url@leostyle{%
  \@ifundefined{selectfont}{\def\UrlFont{}}%
  {\def\UrlFont{}}%
}
\makeatother
\usepackage{xcolor}
\usepackage{booktabs}
\usepackage{dsfont}

\newcommand{\descr}[1]{\smallskip \noindent \textbf{#1}}
\newcommand{\descrit}[1]{\smallskip \noindent \textbf{\em #1}}

\usepackage[all]{xy}
\extrafloats{20}

\newif\ifcomment
\commenttrue 

\makeatletter
\def\subsubsection{\@startsection{subsubsection}{3}%
  \z@{.5\linespacing\@plus.7\linespacing}{.1\linespacing}%
  {\normalfont\itshape}}
\makeatother

\usepackage[marginal,hang]{footmisc}
\renewcommand{\footnoterule}{%
  \kern -3pt
  \hrule width 1in 
  \kern 2pt
}

\renewcommand{\footnotesize}{\fontsize{8}{9}\selectfont}

\ifcomment
	\newcommand{\edc}[1]{\textbf{\em\color{red}[EDC: #1]}}
	\newcommand{\ct}[1]{\textbf{\em\color{blue}[CT: #1]}}
	\newcommand{\ap}[1]{\textbf{\em\color{violet}[AP: #1]}}
	\newcommand{\shep}[1]{\textbf{\em\color{blue}[#1]}}
\else  
    \newcommand\edc[1]{}
    \newcommand\ct[1]{}
    \newcommand\ap[1]{}
    \newcommand\shep[1]{}
\fi

\begin{document}

\title[Measuring Membership Privacy on Aggregate Location Time-Series]{Measuring Membership Privacy on Aggregate Location Time-Series\footnotemark}

\author{Apostolos Pyrgelis}
\email{apostolos.pyrgelis@epfl.ch}
\affiliation{%
  \institution{\'Ecole Polytechnique F\'ed\'erale de Lausanne (EPFL)}
}
\author{Carmela Troncoso}
\email{carmela.troncoso@epfl.ch}
\affiliation{%
  \institution{\'Ecole Polytechnique F\'ed\'erale de Lausanne (EPFL)}
}
\author{Emiliano De Cristofaro}
\email{e.decristofaro@ucl.ac.uk}
\affiliation{%
  \institution{University College London (UCL) \& Alan Turing Institute}
}

\renewcommand*{\thefootnote}{\fnsymbol{footnote}}

\footnotetext{$^*$Presented at ACM SIGMETRICS 2020 and published in the Proceedings of the ACM on Measurement and Analysis of Computing Systems (POMACS), Vol.~2, No.~4, Article 36, June 2020. Work done, in part, while first author was at UCL.}

\makeatletter
\let\@authorsaddresses\@empty
\makeatother

\begin{abstract}
While location data is extremely valuable for various applications, disclosing it prompts serious threats to individuals' privacy. To limit such concerns, organizations often provide analysts with aggregate time-series that indicate, e.g., how many people are in a location at a time interval, rather than raw individual traces. In this paper, we perform a measurement study to understand Membership Inference Attacks (MIAs) on aggregate location time-series, where an adversary tries to infer whether a specific user contributed to the aggregates.
We find that the volume of contributed data, as well as the regularity and particularity of users' mobility patterns, play a crucial role in the attack's success. We experiment with a wide range of defenses based on generalization, hiding, and perturbation, and evaluate their ability to thwart the attack vis-\`a-vis the utility loss they introduce for various mobility analytics tasks.
Our results show that some defenses fail across the board, while others work for specific tasks on aggregate location time-series. For instance, suppressing small counts can be used for ranking hotspots, data generalization for forecasting traffic, hotspot discovery, and map inference, while sampling is effective for location labeling and anomaly detection when the dataset is sparse. Differentially private techniques provide reasonable accuracy only in very specific settings, e.g., discovering hotspots and forecasting their traffic, and more so when using weaker privacy notions like crowd-blending privacy. Overall, our measurements show that there does not exist a unique generic defense that can preserve the utility of the analytics for arbitrary applications, and provide useful insights regarding the disclosure of sanitized aggregate location time-series.
\end{abstract}

\maketitle
\thispagestyle{empty}
\pagestyle{empty}

\renewcommand*{\thefootnote}{\arabic{footnote}}

\section{Introduction}\label{sec:intro}
Location time-series are useful in a wide range of applications, e.g., forecasting traffic volumes and events~\cite{hoang2016forecasting,jiang2018deep}, mining points of interests~\cite{karamshuk2013geo}, monitoring depressive states~\cite{canzian2015trajectories}, etc.
However, mobility patterns expose users' sensitive attributes like lifestyles or religious inclinations, and are hard to anonymize as they inherently contain identifying attributes such as home/work location pairs~\cite{golle2009anonymity}.

Similar to other applications, e.g., smart metering~\cite{smartmet}, financial data~\cite{cfbp} or healthcare~\cite{hu2003interpreting}, analysts often try to avoid these risks by building analytics on top of aggregate data. That is, rather than raw traces, one uses reports of, e.g., how many users are in a certain location during a time interval. %
For instance, Waze builds traffic models using average driving speeds~\cite{waze}, 
Uber provides aggregate data for urban planning purposes~\cite{uber},
and Telefonica monetizes footfall statistics through advertising and business analytics~\cite{smartsteps}.

Even though aggregation is often considered as a privacy-friendly approach to hide sensitive data of single individuals, inference attacks might still be possible on aggregates~\cite{homer2008resolving,wang2009learning,bucher2017}. In the case of aggregate location time-series, an adversary with some prior knowledge about the victims can assess whether or not a specific user contributed to the aggregates; this is known as a {\em Membership Inference Attack} (MIA)~\cite{pyrgelis2018knock} and is the main focus of our study. MIAs on aggregate location time-series have very serious privacy implications: (a) the fact that someone's data is part of the aggregates can be itself sensitive, and (b) these attacks are a first step to other inferences that aim to recover additional information about users such as their mobility profiles~\cite{pyrgelis2017does} and/or their trajectories~\cite{xu2017trajectory} from the aggregates.

In theory, we have a possible solution against membership inference attacks, namely, differentially private statistics~\cite{dwork2010differential}. 
Indeed, publishing statistics under differential privacy bounds the probability of an adversary attempting to distinguish if the data of a particular user was part of them.
However, previous work~\cite{pyrgelis2017does,pyrgelis2018knock} has shown that differentially private mechanisms applied to aggregate location data can only thwart attacks 
at a cost of considerable reductions in accuracy. In other words, they seem to significantly reduce the performance of the attacks only if they introduce a sizeable error in the statistics.

\descr{Understanding MIAs.} To the best of our knowledge, there has not been any study to identify what makes these attacks possible and in what context(s).
More specifically, which spatio-temporal factors contribute to ease inferences and in what scenarios? Which users are most vulnerable, and why?
This lack of understanding hampers our ability to build defenses providing acceptable privacy-utility trade-offs for mobility analytics.
This motivates us to systematically analyze the reasons behind MIA's success, and identify what makes some users more vulnerable than others.
Our analysis %
shows that
the {\em variance} of the location counts (i.e., how many times one has been in that location) over time proves to be the dominant feature.
This inspires us to use Principal Component Analysis (PCA)~\cite{PCA} to let the adversary capture this variability. Besides easing the analysis by reducing dimensionality, PCA also boosts significantly the attack's performance compared to prior work~\cite{pyrgelis2018knock}. 
We then study the importance of various features (e.g., number of events, number of unique locations visited, etc.) for a classifier trained to distinguish users that are most/least vulnerable to PCA-based MIA. Overall, we find that the amount of data contributed to the aggregation has high influence, %
that movements in less popular places/times can reveal a user's presence in the aggregates, and that the attack's success is linked to the uniqueness and regularity of mobility.

\descr{Defending Against MIAs.} Next, we measure the effectiveness of a wide range of defense strategies based on generalization, hiding, and perturbation, which are commonly used in the location privacy literature. %
We observe that there is no single approach, including differentially private techniques, that works well for all tasks.
In fact, %
our aim is to find defenses that provide reasonable utility-privacy trade-offs for specific {\em mobility analytics tasks}.
To some extent, our work follows on the footsteps of other measurement studies like~\cite{jayaraman2019evaluating}, which shows that the guarantees provided by differentially private mechanisms in the context of privacy-preserving machine learning are very pessimistic with respect to the real capabilities of existing attacks. 
Empirically evaluating the effectiveness of defenses is necessary to have a realistic understanding of the risks.
In our study, we measure the privacy gain that defenses yield vis-\`a-vis various utility metrics in the context of a variety of analytics, such as traffic forecasting, hotspot discovery, location labeling, and anomaly detection, computed on aggregate statistics.

Our extensive experiments demonstrate that there is not a single generic defense that mitigates MIA and offers good utility for arbitrary analytics. However, we find that for several specific tasks, there are effective strategies that can retain utility. For instance, suppressing small counts can be used for ranking hotspots, data generalization for forecasting traffic, hotspot discovery, and map inference, while sampling is effective for location labeling and anomaly detection when the dataset is sparse. Differentially private techniques also provide reasonable accuracy, though only in certain settings (e.g., detecting hotspots and forecasting their traffic), and more so when the introduced noise is tailored to achieve weaker notions like crowd-blending privacy.

\descr{Summary of Findings.} 
Overall, our measurement study relies on two real-world datasets representative of different users' mobility patterns and sheds light on the mobility characteristics determining the success of MIA; high mobility and diversity, regularity and uniqueness of moving patterns, as well as location sparsity, are among the factors that contribute to the success of the attack. Moreover, while we do not find a single strategy that defeats MIA in the aggregate setting and that is useful for arbitrary analytics, we identify defenses that provide reasonable tradeoffs for specific mobility analytics tasks.

\descr{Paper Organization.} Next section introduces the datasets and the attacks on which we base our study. We present our feature analysis and the insights it provides in Section~\ref{sec:understanding}. In Section~\ref{sec:defenses}, we evaluate various defense strategies with respect to the protection that they provide against MIA, and, in Section~\ref{sec:priv-util-tradeoff}, we measure their privacy-utility trade-offs for various mobility analytics. Finally, we review related work in Section~\ref{sec:related} and conclude in Section~\ref{sec:conclusion}.

\section{Preliminaries}\label{sec:methodology}
\subsection{Datasets}
Throughout the paper, we perform experiments on two real-world mobility datasets (often used in location privacy research~\cite{shokri2011quantify,pyrgelis2018knock}), which 
represent both regular and irregular mobility patterns disclosed in sporadic and continuous intervals, respectively.

\descr{Transport For London (TFL).} The TFL dataset includes 60M trips made by 4M passengers on the TFL transportation network, between Monday, March 1st and Sunday, March 28th, 2010, using their ``Oyster'' travel cards. For each trip, we have the oyster id (hashed with salt), start time, tap-in station id, end time, and tap-out station id. Trips are made over 582 tube and overground stations, or {\em Regions of Interests (ROIs).} We sample the dataset and retain the data of the 10K oyster ids with the largest amount of trips. We set the time granularity to 1 hour and, for each oyster id, we generate a binary matrix whose rows indicate ROIs and columns indicate timeslots. Each element is 1 if the user tapped-in or out at a certain station, at a certain time, and 0 otherwise. When a passenger does not tap in or out of any station within a given timeslot, we assign it to a special ROI, denoted as \textit{null}. Overall, the sampled dataset contains a total of 7.3M events, with the 10K oysters reporting an average of 728$\pm$16 total tap-in/tap-out events, over 20$\pm$9 unique ROIs, and it is relatively sparse as the oysters are in the transportation system, on average, for 115$\pm$21 out of the 672 hourly slots (28 days).

\descr{San Francisco Cabs (SFC).} The SFC dataset contains mobility trajectories of taxis in the San Francisco area between May 17th and June 10th, 2008~\cite{epfl-mobility-20090224}. Each record includes a cab identifier, latitude, longitude, and a timestamp. Overall, we have 11M GPS coordinates generated by 536 taxis. We retain 3 weeks of data (Monday, May 19th to Sunday, June 8th), keeping only traces within an area of 30.3mi$^2$ to cover downtown San Francisco. We split this area in a uniform grid of 10$\times$10, with cells (ROIs) of 0.3mi$^2$ each, and we set the time granularity to 1 hour. Again, for each cab we generate a binary matrix (with rows representing ROIs and columns timeslots), indicating whether or not a cab transited through a location, at a certain time, and we assign a cab not generating any event within a timeslot to the {\em null} ROI. Overall, the final dataset contains 2M events generated by 534 cabs, each reporting on average 3,827$\pm$1,069 events over 78$\pm$6 unique ROIs; thus, this dataset is much less sparse than TFL, as taxis report ROIs for 340$\pm$94 out of the 504 time slots (21 days).

\descr{Aggregates.} We compute the aggregate location time-series over a group of users (i.e., passengers or cabs) by adding their binary matrices. The resulting matrix is the aggregate time-series and indicates the number of users that transit through the TFL and SFC ROIs, over time.

\begin{table}[t]
\footnotesize
\begin{center}
\setlength{\tabcolsep}{3pt}
\begin{tabular}{ r| l }
\textbf{Symbol} & \textbf{Description} \\
\midrule
Adv, Ch & Adversary, Challenger \\ 
$\mathcal{P}$ & Adversarial Prior Knowledge \\
$\mathcal{U}$ & Set of Mobile Users \\
$\mathcal{S}$ & Set of Locations (ROIs) \\ 
$\mathcal{T}$ & Time Period Considered \\
$\mathcal{T_{O}, T_I}$ & Observation and Inference Periods \\
$m$ & Aggregation Group Size\\
$\alpha$ & Percentage of Users Whose Locations are Known (Subset of Locations Prior) \\
$\beta$ & Number of Groups Whose Aggregates are Known During $\mathcal{T_{O}}$ (Participation in Past Groups Priors) \\
\end{tabular}
\end{center}
\vspace{0.05cm}
\caption{Membership Inference Attacks Notation.}
\label{table:notation}
\vspace{-0.2cm}
\end{table}

\subsection{MIA on Aggregate Location Time-Series}\label{sec:priors}
Our measurement study is based on the attack proposed by Pyrgelis et al.~\cite{pyrgelis2018knock}. In particular, they model membership inference attacks (MIAs) on aggregate location time-series, calculated over a set of locations $\mathcal{S}$ and a set of time intervals $\mathcal{T}$, as a distinguishability game between an adversary (Adv) and a challenger (Ch): Adv needs to distinguish location aggregates, provided by Ch, that include the data of a target user from those that do not. 
The game has several parameters, including the universe of users $\mathcal{U}$, the size of the aggregation groups $m$, and the inference period $\mathcal{T_{I}}$ during which Adv is being challenged. %
Similarly to~\cite{pyrgelis2018knock}, we consider that Adv instantiates her distinguishing function as a classifier trained on data over an observation period $\mathcal{T_{O}}$ available as part of her ``prior knowledge'' $\mathcal{P}$. See Table~\ref{table:notation} for a summary of our notation.

\descr{Priors.} We consider different types of prior knowledge that correspond to different adversaries: %

\begin{enumerate}[topsep=0pt,leftmargin=*]
\item \textbf{Subset of Locations}: Adv knows the actual locations for a subset of users, including her target. The size of the subset is controlled by a parameter $\alpha$. Adversaries with this type of prior knowledge can be telecommunications service providers that collect location data from their users or hackers that compromise providers that store user location data in their databases, and which exploit this information to infer membership of a target user to aggregate datasets published by other entities.

\item Participation in Past Groups: Adv knows her target's participation in aggregates observed during $\mathcal{T_O}\neq\mathcal{T_I}$, with a parameter $\beta$ indicating the number of groups whose aggregates are known. More precisely, we have:
\begin{enumerate}[topsep=0pt]
\item[(2a)] \textbf{Same Groups As Released:} Ch challenges Adv with aggregates computed on the same groups as her prior knowledge; 
\item[(2b)] \textbf{Different Groups Than Released:} Adv is challenged on dynamic groups, which should make the inference harder.
\end{enumerate}
Adversaries with this type of prior knowledge can be location-based services (LBS) that have observed their users' location data over a past period and use that information to infer their membership on aggregate datasets related to future periods. Prior knowledge (2a) corresponds to scenarios of aggregate location data release over stable groups of users, e.g., the user base of a LBS, while (2b) corresponds to scenarios of aggregate release over mixed user groups, e.g., when user data from multiple services such as maps, check-ins, etc., is aggregated.
\end{enumerate}

\descr{Privacy Loss.} We measure the classifier's Area Under the Curve (AUC) score, capturing its performance %
for various classification thresholds, and compute the {\em privacy loss} as the adversarial improvement over a random guess baseline (i.e., AUC score of $0.5$).

\section{Understanding The Attacks}\label{sec:understanding}
In this section, we study {\em what} makes the presence of a user's location data points in aggregates inferable, and {\em how} this varies based on the adversary's prior knowledge. 
\subsection{Experimental Setup}\label{sec:experimental}
As done in~\cite{pyrgelis2018knock}, we split users into three mobility groups ({\em highly, somewhat,} and {\em barely} mobile) and run MIA against 150 users, 50 from each group, sampled at random. To target a user, we create a balanced dataset containing labeled samples of aggregate location time-series that include and exclude her location data to train/test the classifier that is used as a distinguisher by the adversary.

\descr{PCA Optimization.} The methodology proposed in~\cite{pyrgelis2018knock} uses a classifier that takes as input features statistics (i.e., mean, median, maximum, minimum, variance, standard deviation, and sum) computed on every location (ROI) of an aggregate location time-series sample. Initially, we followed this methodology and we performed a standard feature analysis which showed that, in both TFL and SFC datasets, the variance of the location counts over time is among the dominant features. This inspired us to use Principal Components Analysis (PCA) on the target's dataset to reduce the dimensionality of the problem and extract valuable information. PCA converts observations of correlated variables to linearly uncorrelated ones (principal components) via an orthogonal transformation. The first principal component accounts for as much variability in the data as possible, and each succeeding component has the largest variance possible while being orthogonal to the previous components~\cite{PCA}. We feed the resulting components to a Logistic Regression (LR) classifier---we choose LR as it yields the best performance---to infer membership.

\begin{figure}[t]
\centering
\begin{subfigure}{0.4\textwidth}
\includegraphics[width=1.0\textwidth]{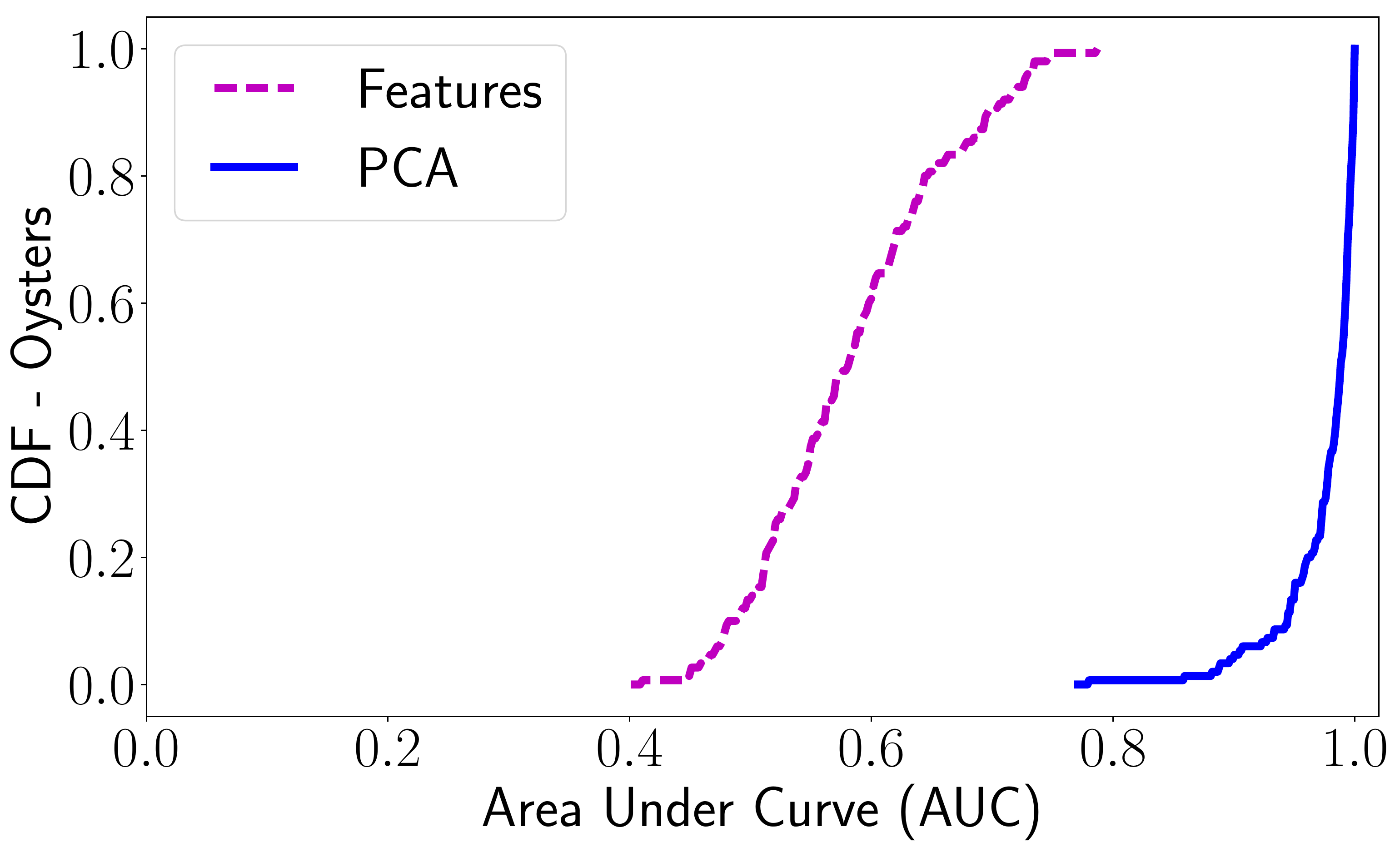}
\caption{TFL ($\alpha{=}0.11$, $m{=}1{,}000$)} %
\end{subfigure}
\hspace{0.2cm}
\begin{subfigure}{0.4\textwidth}
\includegraphics[width=1.0\textwidth]{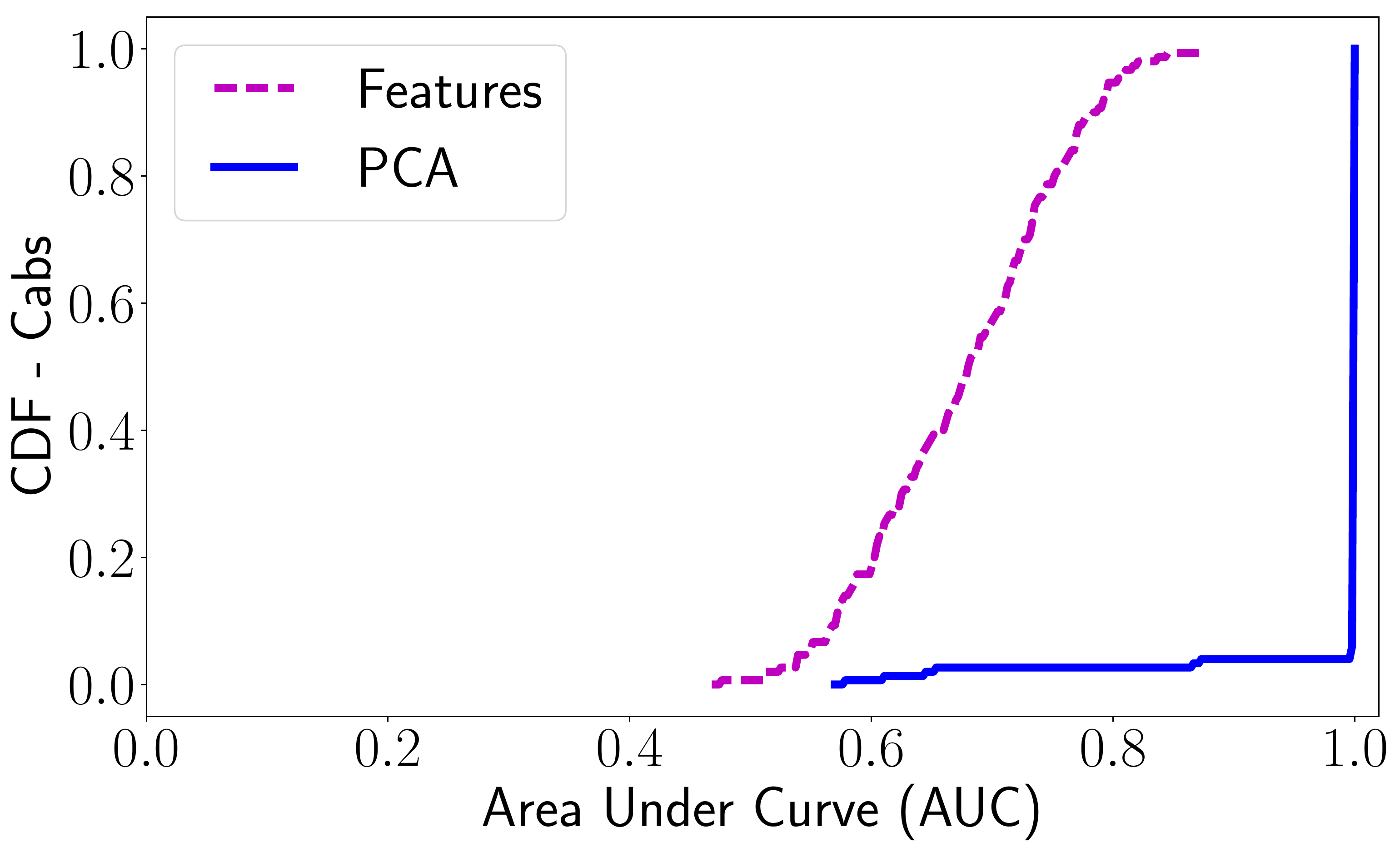}
\caption{SFC ($\alpha{=}0.2$, $m{=}100$)} %
\end{subfigure}
\caption{Performance of membership inference attack when employing the feature extraction  methodology (Features) of~\cite{pyrgelis2018knock} vs. the dimensionality reduction one (PCA) on (a) TFL and (b) SFC datasets. We consider an adversary that has the Subset of Locations Prior and $|\mathcal{T_O}|{=}|\mathcal{T_I}|{=}168$.}
\label{fig:tfl-sfc-feats-vs-PCA}
\vspace{-0.2cm}
\end{figure}

The use of PCA \textbf{boosts the attack's performance while also removing the need for costly feature extraction}. %
To illustrate this improvement, we plot in Figure~\ref{fig:tfl-sfc-feats-vs-PCA} the CDF of the classifier's AUC scores, computed over the 150 target users, for both the feature extraction approach employed in~\cite{pyrgelis2018knock} and the dimensionality reduction one with PCA when using the Subset of Locations type of prior knowledge. The increase on the mean AUC score amounts to 65\% for TFL, and 46\% for SFC. We observe the same trend, though somewhat less prominently, for the other priors: with Same Groups As Released, the improvement is 22\% for TFL and 16\% for SFC; with Different Groups Than Released, it increases by 26\% and 17\%, respectively.

\descr{Analysis Procedure.} To gain a deeper understanding of the attack, we first examine the loadings of the most important (in terms of the LR classifier's weights) principal components for the victims, i.e., their correlation coefficients with the original variables, highlighting which spatial and temporal points contribute to the inference. To gain insights about the differences between victims that are most and least prone to the attack (in terms of AUC score), we train a separate classifier on their mobility characteristics and investigate its features' importance to identify factors making the attack more powerful. Specifically, we compute the following statistics of the users' trajectories: total events (i.e., location-time tuples), unique locations visited, active timeslots, mean locations per timeslot, mean events and active timeslots during week days and weekends, spatial and temporal entropy, and unicity. The latter captures how unique is a user's travel pattern, and is calculated as $unicity_{u}{=} \sum_{t \in \mathcal{T}} \mathds{1}^{t}(u) / | \mathcal{T} |$, where $\mathds{1}^{t}(u)$ indicates if the ordered sequence of locations visited by user $u$ at time $t$ is unique or not. We also consider variables that intuitively can influence the success of MIAs on aggregate locations: volume of data a user contributes, amount of movements in less popular locations/times, and regularity of mobility patterns.

\subsection{Adversaries That Know the Location Data of their Target Users}
\descr{Parameters.} We employ the \textbf{Subset of Locations} prior knowledge and set the percentage of users for which locations are known to the adversary as $\alpha{=}0.11$ and $\alpha{=}0.2$ for the TFL and the SFC experiments, respectively. We consider the maximum user group size that the adversary can attack: $m{=}1{,}000$ for TFL and $m{=}100$ for SFC. In both cases, the first week of data is the observation and inference period (i.e., $| \mathcal{T_O} |{=}| \mathcal{T_I} |{=} 168$ hourly timeslots), and we create datasets of 2K samples to train and test the classifier (with a 80$\%$--20$\%$ train/test split). We extract 1K principal components for TFL and 600 for SFC (which account for more than $99\%$ of the variance in the victims' datasets, in each case).

\begin{figure*}[t]
\centering
\begin{subfigure}{0.3\textwidth}
\includegraphics[width=1.0\textwidth]{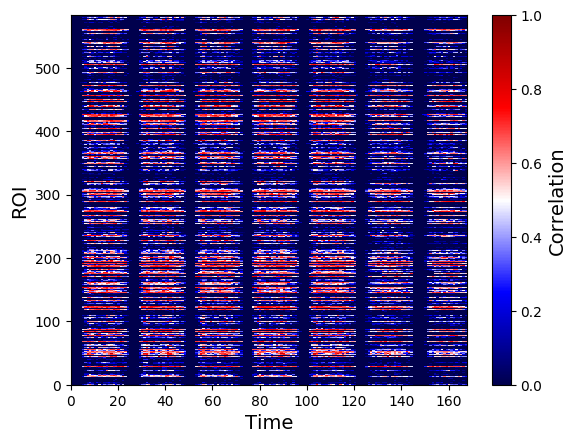}
\caption{Original Heatmap}
\label{fig:tfl-agg-10perc}
\end{subfigure}
\hspace{0.2cm}
\begin{subfigure}{0.3\textwidth}
\includegraphics[width=1.0\textwidth]{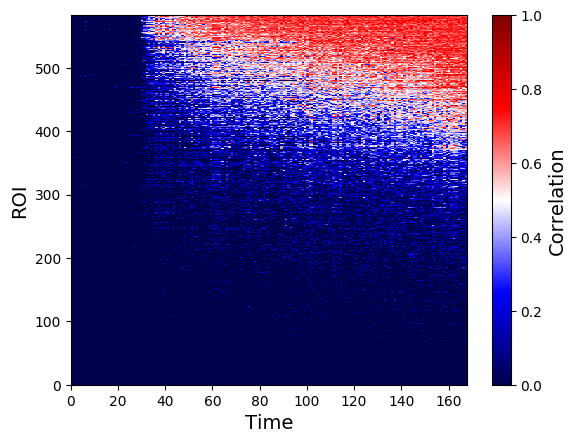}
\caption{All Victims (Sorted)}
\label{fig:tfl-agg-loc-time-10perc}
\end{subfigure}
\\
\begin{subfigure}{0.3\textwidth}
\includegraphics[width=1.0\textwidth]{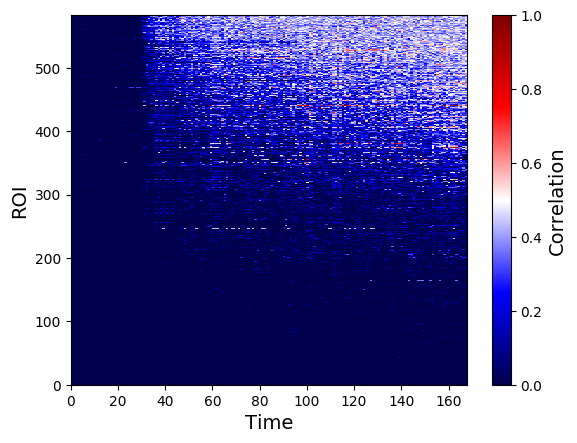}
\caption{Top 10\% Victims (Sorted)}
\label{fig:tfl-agg-loc-time-10perc-top}
\end{subfigure}
\hspace{0.2cm}
\begin{subfigure}{0.3\textwidth}
\includegraphics[width=1.0\textwidth]{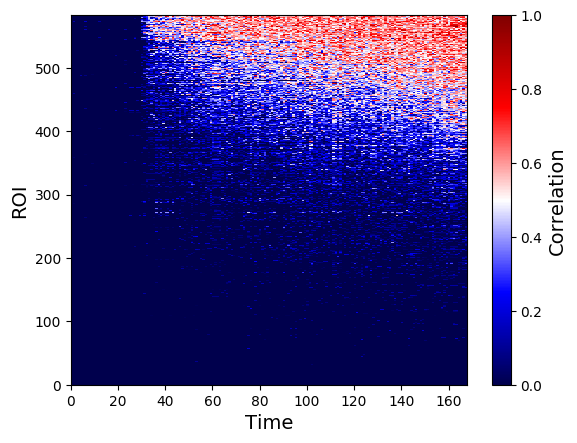}
\caption{Bottom 10\% Victims (Sorted)}
\label{fig:tfl-agg-loc-time-10perc-bottom}
\end{subfigure}
\caption{Subset of Locations Prior, TFL ($\alpha{=}0.11$, $m{=}1,000$, $|\mathcal{T_O}|{=}|\mathcal{T_I}|{=}168$). Aggregate spatio-temporal correlation over the 2 most important components per victim: (a) original heatmap; heatmap sorted (ascending-order) by location and timeslot popularity computed on (b) all victims, (c) top 10$\%$, and (d) bottom 10$\%$ of distinguishable victims.}
\vspace{-0.2cm}
\end{figure*}

\descr{Correlation Coefficients.} For TFL, in Figure~\ref{fig:tfl-agg-10perc}, we plot a heatmap of the components' correlation with the original spatio-temporal points in the data, aggregated (and normalized) over the 2 most important principal components of each victim. We observe how events in various locations and times yield high correlation values (dark red), i.e., {\em diverse events contribute to membership inference}. First, we observe that different stations exhibit different levels of correlation (possibly due to their location, e.g., central vs. residential ones). Second, we see differences between the patterns of week days (ids 1--120) and of weekends (121--168), with {\em commuting hours having high correlation values}. Interestingly, for some ROIs, busy mid-day hours also yield high correlation. The same happens with weekend events (right side of the heatmap): users' presence in the aggregates might be revealed if they travel at times when the transportation system is less crowded.

\begin{figure*}[t]
\centering
\begin{subfigure}{0.3\textwidth}
\includegraphics[width=1.0\textwidth]{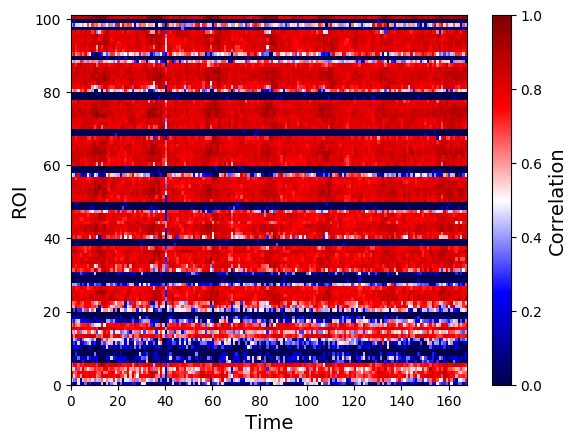}
\caption{Original Heatmap}
\label{fig:sfc-agg-20perc}
\end{subfigure}
\hspace{0.2cm}
\begin{subfigure}{0.3\textwidth}
\includegraphics[width=1.0\textwidth]{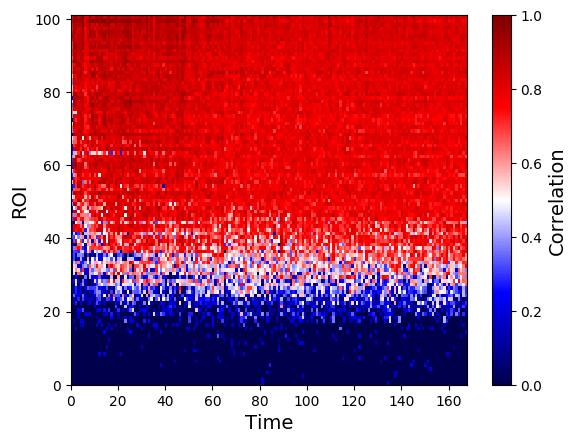}
\caption{All Victims (Sorted)}
\label{fig:sfc-agg-loc-time-20perc}
\end{subfigure}
\\
\begin{subfigure}{0.3\textwidth}
\includegraphics[width=1.0\textwidth]{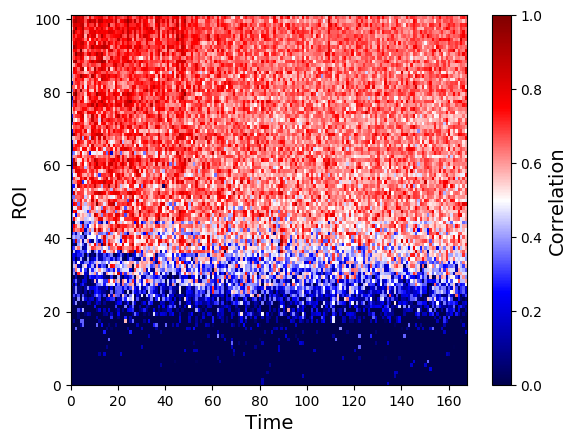}
\caption{Top 10\% Victims (Sorted)}
\label{fig:sfc-agg-loc-time-20perc-top}
\end{subfigure}
\hspace{0.2cm}
\begin{subfigure}{0.3\textwidth}
\includegraphics[width=1.0\textwidth]{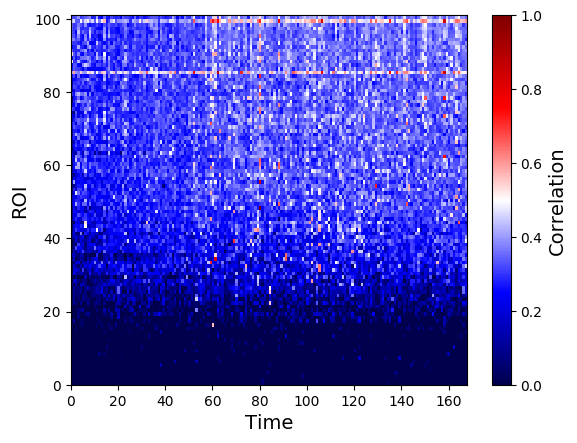}
\caption{Bottom 10\% Victims (Sorted)}
\label{fig:sfc-agg-loc-time-20perc-bottom}
\end{subfigure}
\caption{Subset of Locations Prior, SFC ($\alpha{=}0.2$, $m{=}100$, $|\mathcal{T_O}|{=}|\mathcal{T_I}|{=}168$). Aggregate spatio-temporal correlation over the 5 most important components per victim: (a) original heatmap; ascending-order sorted heatmap by location and timeslot popularity computed on (b) all victims, (c) top 10$\%$, and (d) bottom 10$\%$ of distinguishable victims.}
\vspace{-0.2cm}
\end{figure*}

For SFC, in Figure~\ref{fig:sfc-agg-20perc}, we plot a heatmap of the aggregate (and normalized) spatio-temporal correlation coefficients, computed over the cabs' 5 most important components. As opposed to TFL, {\em a large number of locations yield high coefficients, highlighting that GPS movements offer a larger attack surface} than tap-in/out events at London stations. We see a similar effect for time, with {\em high correlations in mid-day hours}, but also during weekends when fewer drivers are working.

\descr{Location/Timeslot Popularity.} %
In Figure~\ref{fig:tfl-agg-loc-time-10perc}, we sort the TFL aggregate correlation heatmap, both locations and time, according to their frequency of appearance in the dataset. As expected, the {\em more popular locations/timeslots yield the highest correlation} (upper-right corner), since most of the events are generated in such locations/times. Nonetheless, data points in popular locations but reported in less popular timeslots (upper-left corner) are also important, suggesting that {\em commuters can be distinguished in the aggregates if they visit such locations at non-busy times}. Finally, a few points in less popular locations (and various times) have high coefficients, i.e., {\em movements in sparse locations/times can give away a commuter's presence in the aggregates}.

Similarly, for the SFC dataset, we find that the most popular locations yield high coefficients; see Figure~\ref{fig:sfc-agg-loc-time-20perc}. However, contrary to TFL where only a small subset of stations and times are relevant (recall Figure~\ref{fig:tfl-agg-loc-time-10perc}), in SFC also mid-popular locations (i.e., ids 40-60), as well as certain hours in less popular ROIs (ids 20-40), obtain significant correlation. This suggests that popularity is not as important as in TFL, {\em there are many regions and times that help membership inference}.

\descr{Susceptibility to MIA.} We then analyze the features of the most and least distinguishable victims of MIA. For the top distinguishable victims of the TFL dataset (Figure~\ref{fig:tfl-agg-loc-time-10perc-top}), we find very high coefficients for relatively unpopular locations and times (middle part of the heatmap), i.e., {\em people visiting rare locations at rare times are easy to attack}. The most popular places and times (top right) do yield high correlation, but they do not seem to be as crucial. For the less distinguishable commuters (Figure~\ref{fig:tfl-agg-loc-time-10perc-bottom}), popular locations and times (top right part of the heatmap) are the most important, and no other locations seem to help the attack. We believe that these are commuters that mostly travel at popular stations/times and their data hides better along with those of the crowd. 

For the SFC dataset, the top distinguishable cabs (Figure~\ref{fig:sfc-agg-loc-time-20perc-top}) yield high coefficients overall and slightly higher ones in the most popular locations but during the least busy times. This confirms that {\em movements in less popular times enhance membership inference}. For these cabs, the counts of popular locations and timeslots also yield high correlation values, i.e., they contribute a large portion of points during the inference week. The heatmap of the least distinguishable cabs (Figure~\ref{fig:sfc-agg-loc-time-20perc-bottom}) is much sparser: most cabs contribute little data and the attack has little information to build on. Similarly to TFL, the most popular locations and timeslots, where and when most cabs contribute to anyway, yield the highest correlation, i.e., popular regions and times are not very revealing.

\begin{figure}[t]
\centering
\begin{subfigure}{0.4\textwidth}
\includegraphics[width=1.0\textwidth]{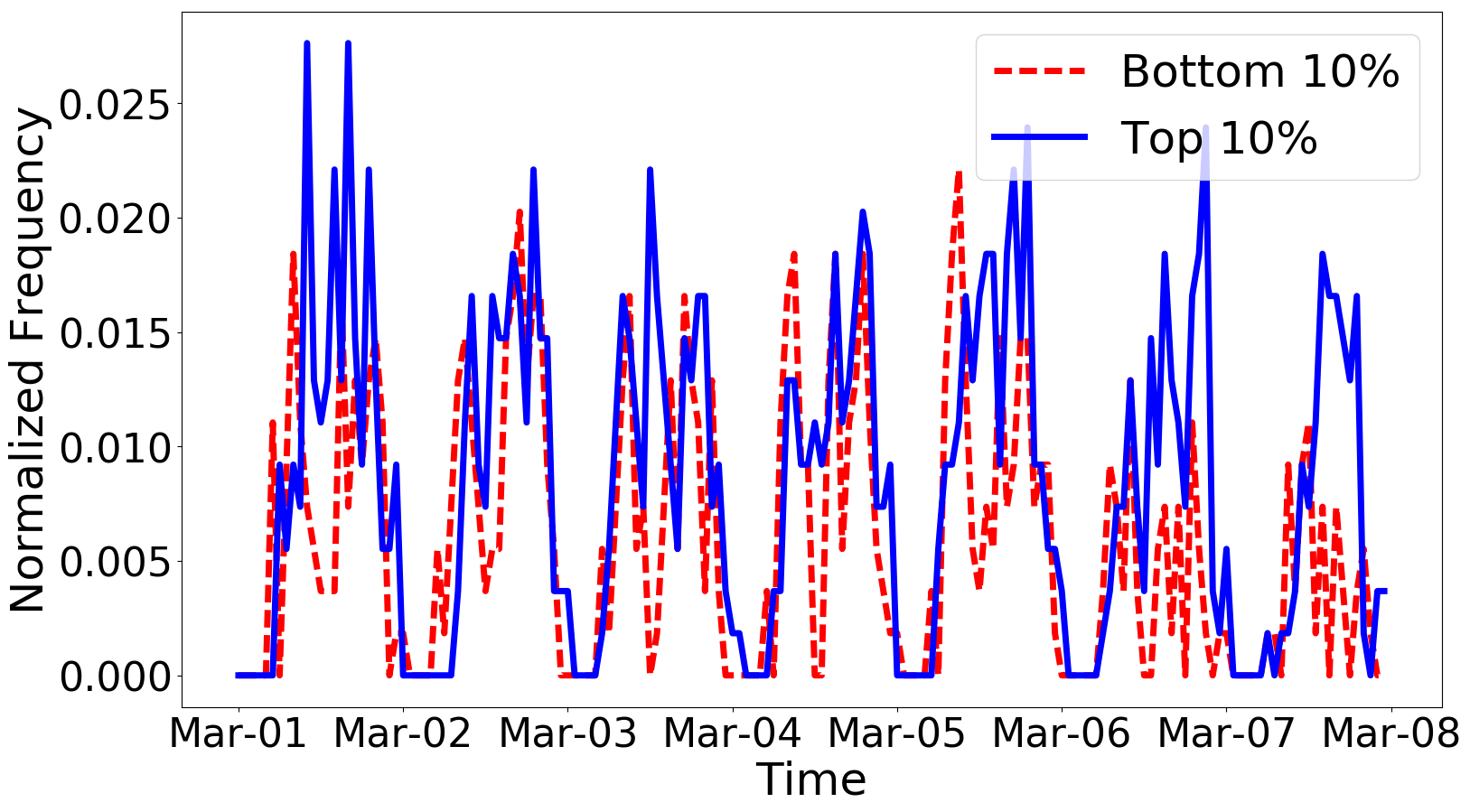}
\caption{TFL ($m{=}1,000$, $|\mathcal{T_I}|{=}168$).}
\label{fig:tfl-time}
\end{subfigure}
\hspace{0.2cm}
\begin{subfigure}{0.4\textwidth}
\includegraphics[width=1.0\textwidth]{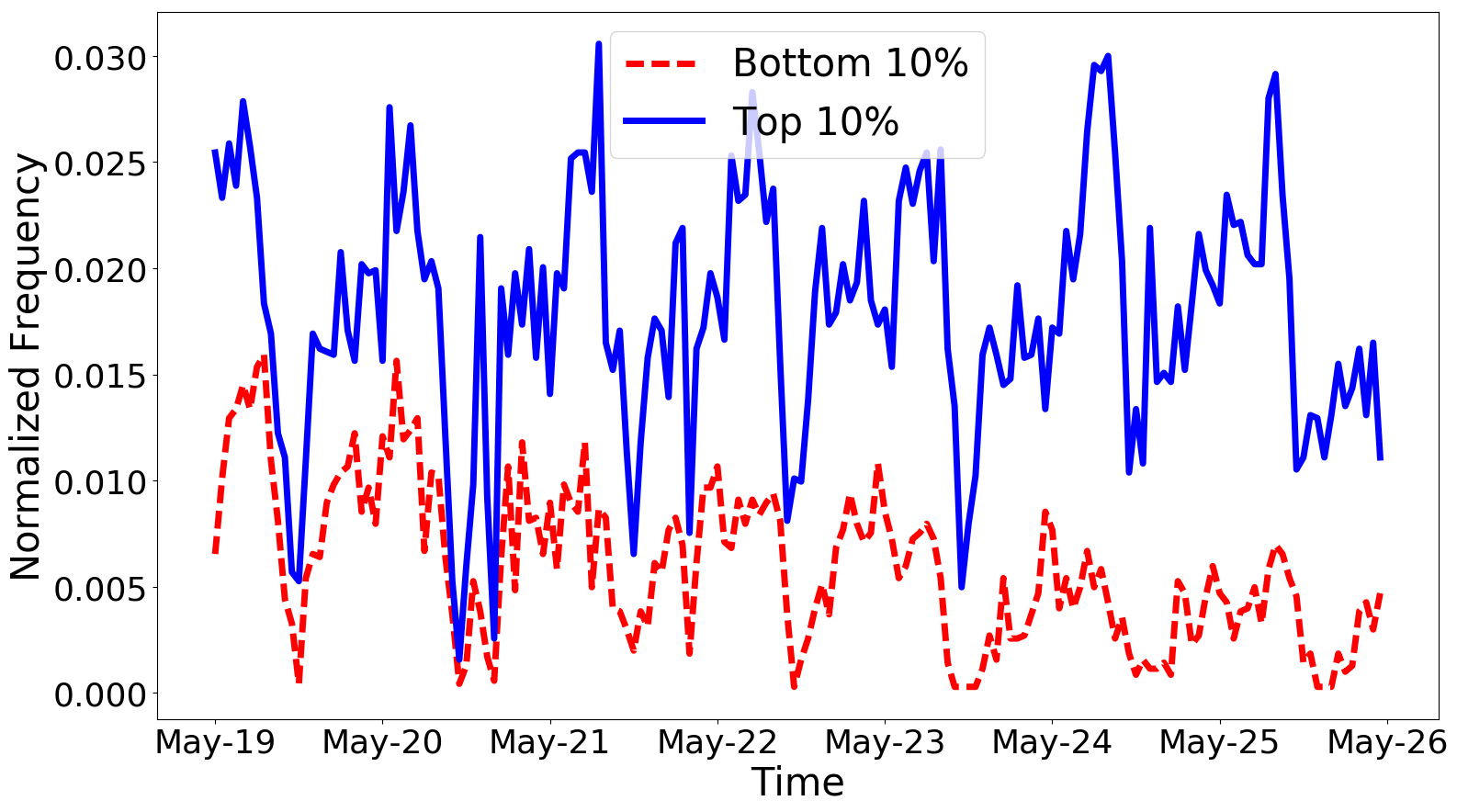}
\caption{SFC ($m{=}100$, $|\mathcal{T_I}|{=}168$).}
\label{fig:sfc-time}
\end{subfigure}
\caption{Normalized frequency of top/bottom 10\% distinguishable victims over $\mathcal{T_I}$ for (a) TFL and (b) SFC datasets.}
\label{fig:tfl-sfc-time}
\vspace{-0.2cm}
\end{figure}

To verify the above hypotheses, for both datasets, we study the aggregate (normalized) frequency of the timeslots in the inference week over the two groups in Figure~\ref{fig:tfl-sfc-time}. For TFL (Figure~\ref{fig:tfl-time}), the distinguishable commuters (blue line) have higher frequency in mid and late evening hours of weekdays and during weekends; i.e., {\em sporadic movements when the transportation system is less crowded facilitate membership inference}. The less distinguishable oysters (red line) mostly move during ``commuting'' hours thus they are hard to pick apart from other commuters. For SFC, Figure~\ref{fig:sfc-time} shows that the top distinguishable cabs (blue line) have higher frequencies in the late night hours on weekdays and during weekends, i.e., {\em movements in low activity hours facilitate attacks}. Whereas, the less distinguishable cabs (red line) contribute some data at the beginning of the week, but less afterwards. This confirms that the most distinguishable cabs are those which contribute larger volume of spatio-temporal points, i.e., {\em bigger data contribution enhances MIA performance.}

Finally, for both datasets, we feed the mobility characteristics (see Section~\ref{sec:experimental}) of the top and bottom 10$\%$ distinguishable oysters to a Random Forest classifier and examine which features can separate the two groups (see Table~\ref{table:feats-tfl-sfc}). For TFL, the most important feature is the number of unique locations visited by a user: {\em visiting more (unique) locations increases the attack's surface and thus its success}. Second in importance is the unicity, highlighting a link between MIA and the uniqueness of mobility patterns: top victims have a unique mobility pattern for 14$\pm$5 timeslots of $\mathcal{T_I}$, while the bottom ones are unique only for 4$\pm$1. For SFC (again, see Table~\ref{table:feats-tfl-sfc}), the top feature is the mean number of locations per timeslot followed by the number of active timeslots and total number of events, showing that unlike in TFL where most commuters report similar volumes of data, in the SFC dataset {\em vehicles with more data points are overall more susceptible to MIAs}. We also observe that, similarly to TFL, MIA's performance is strongly linked to the uniqueness of cabs' mobility trajectories. The top distinguishable cabs exhibit larger unicity (their patterns are unique for 124$\pm$6 out of the 168 timeslots) than the bottom ones (unique pattern for 38$\pm$30 timeslots).

\begin{table}[t]
\footnotesize
\begin{center}
\setlength{\tabcolsep}{3pt}
\begin{tabular}{ lrr|lrr }
\textbf{Feature} & \textbf{TFL} & \textbf{SFC} & \textbf{Feature} & \textbf{TFL} & \textbf{SFC} \\
\midrule
Total events & 0.03 & 0.17 & Events/weekday & 0.01 & 0.07\\
Unique locations & {\bf 0.39} & 0.01 & Events/weekend & 0.13 & 0.03\\
Active timeslots & 0.06 & 0.23 & Spatial entropy & 0.01 & 0.03\\
Locations per timeslot & 0.05 & {\bf 0.30} & Temporal entropy & 0.06 & 0.01\\
Active timeslots/weekday & 0.01 & 0.01 & Unicity & 0.16 & 0.17\\
Active timeslots/weekend & 0.11 & 0.01\\
\end{tabular}
\end{center}
\vspace{0.05cm}
\caption{Subset of Locations Prior. Feature importance for a Random Forest classifier separating top/bottom 10$\%$ distinguishable victims: TFL ($\alpha{=}0.11$, $m{=}1,000$, $|\mathcal{T_O}|{=}|\mathcal{T_I}|{=}168$) and SFC ($\alpha{=}0.2$, $m{=}100$, $|\mathcal{T_O}|{=}|\mathcal{T_I}|{=}168$).}
\label{table:feats-tfl-sfc}
\vspace{-0.2cm}
\end{table}

\subsection{Adversaries That Know the Target User's Participation in Aggregates Related to the Past}

Next, we analyze the case where Adv knows the target victim's participation in aggregate locations released during an observation period $\mathcal{T_O}\neq\mathcal{T_I}$. We consider two settings for this prior: (a) \textbf{Same Groups As Released}, where the Adv performs inference on the same groups as in the observation period, and (b) \textbf{Different Groups Than Released}, where the inference is made on different groups.

\descr{Parameters.} For both sets of experiments, we set the size of the groups to $m{=}9{,}500$ for TFL and $m{=}500$ for SFC. We consider $\mathcal{T_O}$ to be the first weeks of each dataset (i.e., $| \mathcal{T_O} |{=}504$ for TFL and $| \mathcal{T_O} |{=}336$ for SFC), and use them to construct the prior knowledge the adversary relies on to train her classifier. The attack is run on the last week of the data (i.e., $| \mathcal{T_I} |{=}168$). We configure the number of known groups as $\beta{=}500$ and $\beta{=}800$, for TFL and SFC, creating large enough training/testing datasets---of 2K (2.4K) samples  for TFL (SFC). Moreover, we keep 1K (600) principal components for TFL (SFC), which explain more than 99\% of the variance in each victim's dataset.

\begin{figure}[t]
\centering
\hspace*{-0.15cm}
\begin{subfigure}{0.3\textwidth}
\includegraphics[width=1.0\textwidth]{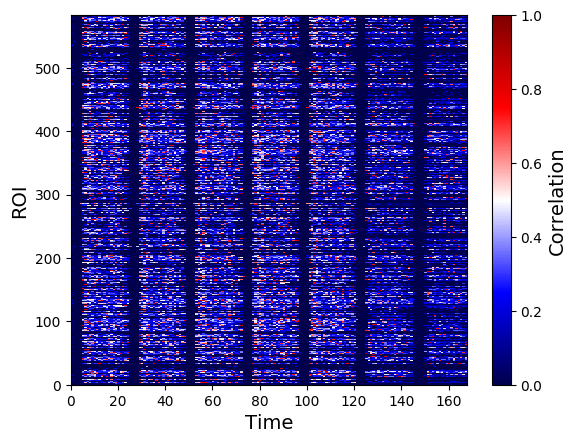}
\caption{TFL}
\end{subfigure}
\hspace{0.2cm}
\begin{subfigure}{0.3\textwidth}
\includegraphics[width=1.0\textwidth]{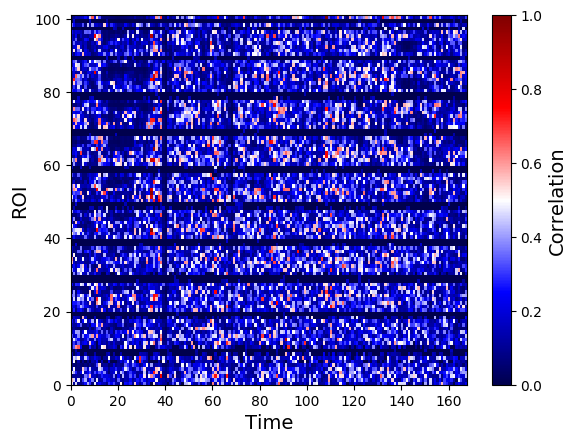}
\caption{SFC}
\end{subfigure}

\caption{Same Groups As Released Prior. Aggregate spatio-temporal correlation over the most important components per victim for (a) TFL ($\beta{=}500$, $m{=}9{,}500$, $|\mathcal{T_O}|{=}504$, $|\mathcal{T_I}|{=}168$) and (b) SFC ($\beta{=}800$, $m{=}500$, $|\mathcal{T_O}|{=}336$, $|\mathcal{T_I}|{=}168$).}
\label{fig:tfl-sfc-agg-same-groups}
\vspace{-0.2cm}
\end{figure}

\descr{Correlation Coefficients.} We start with the Same Groups As Released setting, where Adv performs MIA on the same groups as those on which she trained her classifier. We plot the aggregate correlation coefficients for the most important components of the victims in TFL (top 1 component/victim) and SFC (top 5 components/victim) in Figure~\ref{fig:tfl-sfc-agg-same-groups}. For TFL, the most correlated data points now occur during the morning commuting hours of weekdays, highlighting that \emph{regularity in mobility patterns, e.g., the daily commute to work, helps MIA}. This explains the MIA's great success on TFL with this type of prior~\cite{pyrgelis2018knock}. Interestingly, we also find that popular locations/times as well as less popular locations on popular times (and vice versa) contribute to the inference, showing that commuters exhibit different regular patterns that are equally useful for the attack.

For SFC, movements in some weekdays' slots yield high correlation, i.e., there are some regular cabs that are more susceptible to MIA than others. Looking at the location and timeslot popularity, we find high correlations scattered in the spatio-temporal space. Nonetheless, we observe that movements during less popular timeslots obtain slightly higher correlation values in the components, i.e., cabs regular at such times are prone to MIA. Overall, {\em the attack does not work very well with this prior when victims do not exhibit the same mobility patterns over the weeks,} as most cabs in SFC. 

Our above insights also hold for the Different Groups Than Released setting, for both datasets: regular mobility patterns contribute to the success of MIA. Nevertheless, it is not clear {\em what} locations or times are more important, i.e., various types of regular patterns make MIA successful.

\descr{Susceptibility to MIA.} With the Same Groups As Released prior, all TFL commuters are harmed equally so no analysis is needed. For SFC, the insights are similar to those for the Different Groups Than Released prior, discussed next. In particular, for the Different Groups Than Released setting, we compare the mobility characteristics of the top and bottom 10$\%$ distinguishable victims using a Logistic Regression classifier. Table~\ref{table:feats-tfl-sfc-diff-groups} reports the model's coefficients for each feature (negative and positive coefficients indicate the more and less distinguishable victims, respectively).

For both TFL and SFC, the strongest feature for the top victims is the uniqueness of mobility patterns, i.e., {\em the more unique movements are, the easier it is to infer membership on dynamic groups}. With TFL, we find that the top victims have unique pattern for 47$\pm$13 out of the 672 hourly timeslots, while the bottom ones exhibit unicity for 32$\pm$7 timeslots. 
With SFC, top victims exhibit unique mobility for 357$\pm$45 out of the 504 timeslots, and the bottom ones are unique for 287$\pm$85 timeslots. Features related to time patterns also play an important role in separating the two groups for TFL; e.g., the top distinguishable users are mostly contributing events during weekdays, the bottom ones in the weekends. The results also suggest that top users are mostly regular weekday commuters in less popular ROIs (and thus more unique), while bottom ones travel to locations outside their ``regular'' pattern during weekends. This is confirmed by other features with high coefficients, e.g., the number of unique locations and spatial entropy. For SFC, the amount of data contributed yields stronger features for the more distinguishable cabs, i.e., `regular' cabs reporting larger volumes of data are more identifiable. For the least distinguishable cabs, the number of unique locations is stronger. Overall, this confirms that when the adversary trains on past groups, {\em showing up in many locations, but without repeating patterns, can reduce her power}.

\begin{table}[t]
\begin{center}
\footnotesize
\setlength{\tabcolsep}{3pt}
\begin{tabular}{ lrr|lrr }
\textbf{Feature} & \textbf{TFL} & \textbf{SFC} & \textbf{Feature} & \textbf{TFL} & \textbf{SFC} \\
\midrule
Total events & 0.20 & -0.36 & Events/weekday & -0.47 & -0.38\\
Unique locations & {\bf 0.78} & {\bf 1.29} & Events/weekend & 0.64 & -0.17\\
Active timeslots & -0.17 & 0.05 & Spatial entropy & 0.52 & -0.18\\
Locations per timeslot & 0.01 & -0.33 & Temporal entropy & 0.17 & -0.06\\
Active timeslots/weekday & -0.48 & 0.28 & Unicity & {\bf -1.55} & {\bf -0.68}\\
Active timeslots/weekend & 0.42 & -0.46\\
\end{tabular}
\end{center}
\vspace{0.05cm}
\caption{Different Groups Than Released Prior. Model coefficients of a Logistic Regression classifier separating top/bottom 10$\%$ distinguishable victims: TFL ($\beta{=}500$, $m{=}9,500$, $|\mathcal{T_O}|{=}504$, $|\mathcal{T_I}|{=}168$) and SFC ($\beta{=}800$, $m{=}500$, $|\mathcal{T_O}|{=}336$, $|\mathcal{T_I}|{=}168$).}
\label{table:feats-tfl-sfc-diff-groups}
\vspace{-0.2cm}
\end{table}

\subsection{Take Aways}

Our analysis provides several interesting insights. First, it shows that the performance of the attacks can be significantly boosted using dimensionality reduction techniques such as PCA. More specifically, we get up to 65\% increase in the mean AUC for TFL and 46\% for SFC compared to previous work~\cite{pyrgelis2018knock}. This is because aggregate location data retains strong spatio-temporal correlations with the data provided by individual users. 

Second, we also find the spatio-temporal correlations within the principal components to be aligned with the mobility patterns in the data. For instance, commuting patterns emerge quite clearly in the components of the TFL dataset, while dense GPS trajectories create a large attack surface to be exploited in SFC. In both cases, there is a variety of spatio-temporal points and trends that contribute towards to the inference's success. This makes challenging to design generic and robust defenses that protect \emph{every} user. 

Third, we identify the factors that affect the success of the inference. We show that: 1) Users who contribute more data points to the aggregation are more susceptible to MIA; 2) Movements in sparse locations/timeslots can give away one's presence in the aggregates; 3) Unique mobility patterns are identifiable in the aggregates when the adversary knows about them; and 4) Regular mobility patterns can reveal the users' contribution to the aggregates, even if the adversary observes these patterns in the past.

Finally, we discover factors that {\em negatively} affect the attack's performance; e.g., presence in popular locations and times generally limits inferences, and so do irregular movements that do not repeat over time, in particular, when the adversarial prior knowledge is built on information related to the users' past.

\section{Defenses}\label{sec:defenses}

In this section, we measure the effectiveness of various defense strategies against MIA. We explore a wide range of strategies commonly used in the location privacy literature -- namely, generalization, hiding, and perturbation -- adapted to the aggregate setting with the goal of hiding the features that give membership away according to the analysis in the previous section.

\subsection{Preliminaries}
\descr{Experimental Setup.} We focus on the cases where MIA works best, to evaluate a {\em worst case scenario} for the defenses. For TFL, we consider the Same Groups As Released prior knowledge, with $\beta{=}500$, groups of $m{=}9{,}500$ users, and $\mathcal{T_O}$ being the first 3 weeks of the dataset and $\mathcal{T_I}$ the last one (i.e., $|\mathcal{T_O}|{=}504$ and $|\mathcal{T_I}|{=}168$). For SFC, we choose the Subset of Locations prior, with $\alpha{=}0.5$, $m{=}250$, and both $\mathcal{T_O}$ and $\mathcal{T_{I}}$ being the first week (i.e., $|\mathcal{T_O}|{=}|\mathcal{T_I}|{=}168$). We create large enough datasets of 2K samples for TFL and 2.4K for SFC, and we extract 1K and 600 principal components, respectively. We also consider a strategic adversary that knows the mechanism employed by the defender and can use this to optimize her training---i.e., the adversary trains the classifier on aggregates perturbed using that defense strategy.

\descr{Privacy Gain (PG) Metric.} Following~\cite{pyrgelis2018knock}, we measure the effectiveness of the defenses as the normalized decrease in the adversarial performance:%
~
\begin{equation} \label{eq:pg}
\text{PG} = \begin{cases}
\frac{\text{AUC}_{A} - \text{AUC}_{A'}}{\text{AUC}_{A} - 0.5} & ~~~~~\text{if} ~ \text{AUC}_{A} > \text{AUC}_{A'} \geq 0.5  \\
 0 & ~~~~~ \text{otherwise} \\
\end{cases}%
\end{equation}
where $\text{AUC}_{A}$ is the attack's performance on raw aggregates, and $\text{AUC}_{A'}$ is its performance {\em after} a defense has been applied. PG is a value between 0 and 1, capturing how much the inference power drops towards the random guess baseline (AUC score of 0.5) where users have \textit{perfect} privacy.
\subsection{Experimental Evaluation}\label{sec:def}

\begin{figure*}[t]
\centering
\begin{subfigure}{0.8\textwidth}
 \includegraphics[width=1\textwidth]{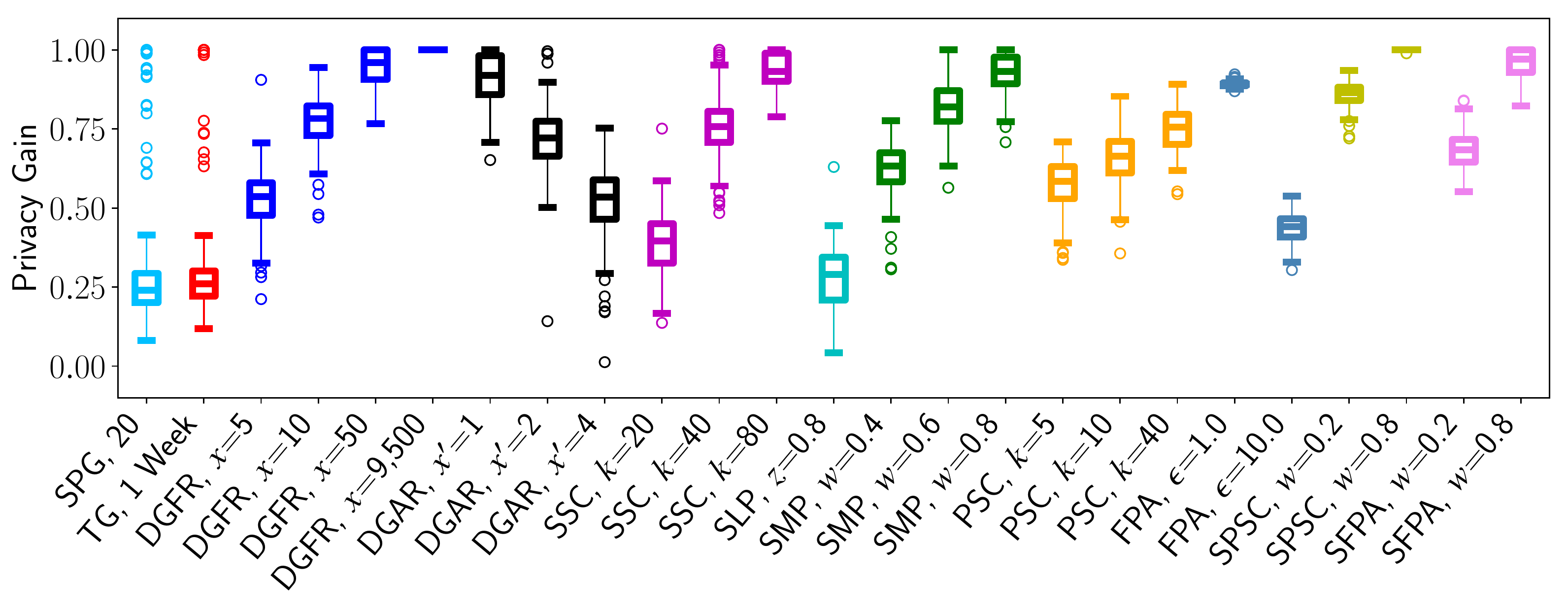}
 \caption{TFL}
 \label{fig:tfl-defenses}
\end{subfigure}
\begin{subfigure}{0.8\textwidth}
\includegraphics[width=1\textwidth]{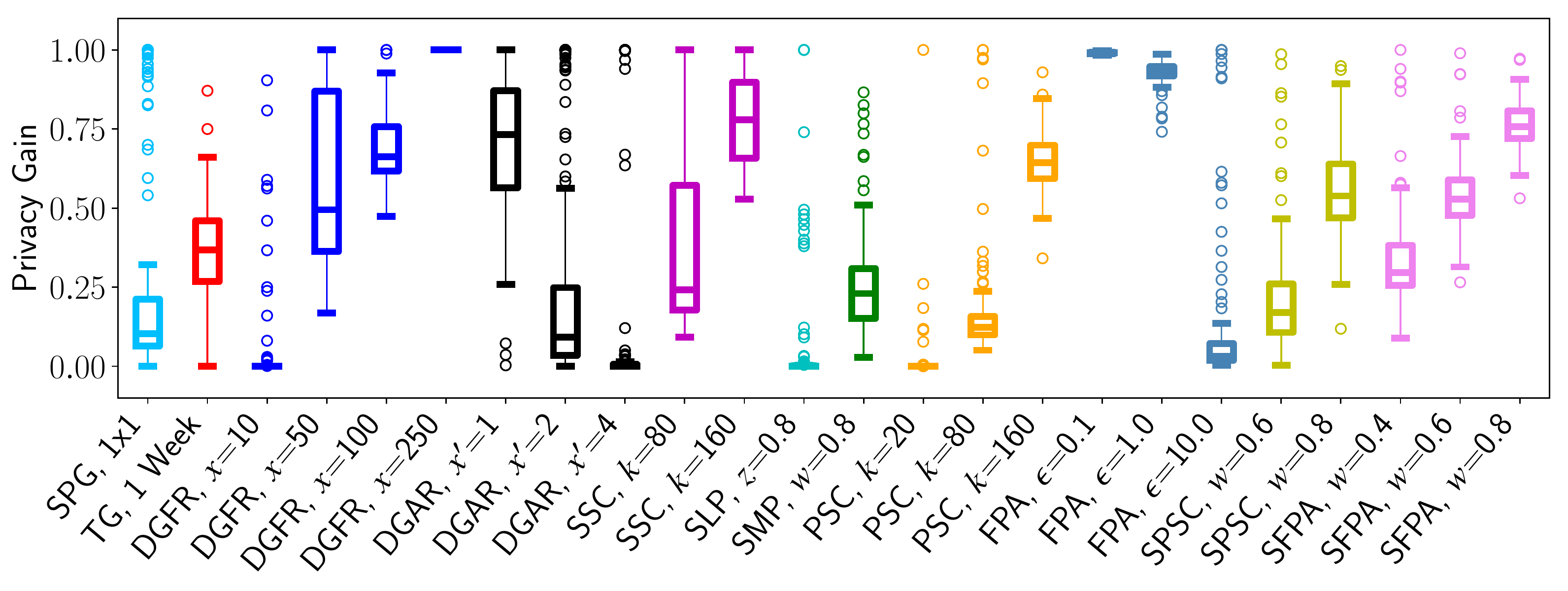}
\caption{SFC}
\label{fig:sfc-defenses}
\end{subfigure}
\caption{Privacy Gain for Various Defenses \& Parameters.}
\end{figure*}

To ease presentation, we use boxplots reporting the privacy gain of the users that we attack for the most interesting defenses configurations; see Figures~\ref{fig:tfl-defenses} for TFL and~\ref{fig:sfc-defenses} for SFC. In the rest of this section, we dive deeper into the discussion of the results.

\subsubsection{Generalization}\label{sec:generalization}

Generalization reduces the precision with which spatio-temporal events are reported~\cite{GruteserG03,krumm2007inference}, and thus their uniqueness~\cite{de2013unique}. 
Reporting data in ranges rather than releasing exact statistics, e.g., using bucketing techniques, is also a generalization technique. 
It has been used to protect privacy in domains such as website fingerprinting, by obfuscating the length of network packets~\cite{cai2014systematic}; or social network privacy, by providing inexact statistics to advertisers~\cite{venkatadri2018privacy}.

\descr{Spatial Generalization (SPG).} We first experiment varying the spatial granularity, while keeping the original temporal resolution (1 hour). This technique has been used to decrease the uniqueness of mobility traces~\cite{de2013unique}, which our analysis showed to be correlated with the success of MIA. For TFL, we group nearby stations -- with the group size being a parameter which we configure in the set \{5, 10, 20\} -- and compute their combined aggregates. We find that only when 20 stations are grouped together, there is a small increase in PG (0.30 on average), with few outliers reaching higher protection. 

For SFC, we use grids of different spatial resolution to divide up the 30mi$^2$ area of downtown San Francisco, ranging from a baseline 10$\times$10 grid resulting in ROIs of 0.3mi$^2$, to one single ROI of 30.3mi$^2$ (with the intermediate grid sizes being 5$\times$5 and 2$\times$2). Only when we consider a single ROI (1$\times$1 grid) the PG increases slightly (0.25 on average), nonetheless, PG$\leq$0.23 for 75\% of the cabs. This means that the temporal dimension of the location contains enough information for the attack to succeed, when the adversary has the `Subset of Locations' prior. In other words, such an attacker can perform MIA on the SFC dataset \emph{even without any spatial information}.

\descr{Temporal Generalization (TG).} We then vary the length of the timeslots from 1 hour (the baseline) to 1 week, %
keeping the original spatial resolution. For TFL, MIA's performance only decreases significantly when the slots aggregate information for one week. Still, PG increases only to 0.31 on average with few outliers. This means that {\em regular commuting patterns remain distinguishable in the aggregates even for relatively long periods of time}. In SFC, the attack's performance starts degrading earlier (i.e., PG${=}$0.15 for 1 day resolution), and while PG reaches 0.35 with a 1-week timeslot, it remains less than 0.27 for 25\% of the cabs. This means that just 1 time point may be sufficient for the attack to succeed. Overall, even without the temporal dimension, the spatial domain still contains enough information to perform inference.

In theory, if one simultaneously generalized both space and time, MIA would be successfully mitigated; e.g., for 1$\times$1 grid with 1-week temporal resolution, PG reaches 0.96 for SFC. However, in such a setting, the aggregates are not useful at all (see Section~\ref{sec:priv-util-tradeoff}).

\descr{Data Generalization (DG).} Finally, we experiment with releasing {\em ranges}: for instance, we report the range `120--130' when 124 users are in a given ROI during a 1-hour timeslot.

\descrit{Data Generalization with Fixed Ranges (DGFR).} We first experiment with a fixed range size (denoted by $x$) for all ROIs, configuring it from $x{=}2$ up to the maximum possible value of the aggregates (as indicated by $m$). To generalize the location counts we assign them to the median of the corresponding range. For TFL, we consider $x$ in the set \{2, 5, 10, 50, 150, 9,500\}. We observe a significant gain for $x{\in}$ \{5, 10\}, with 50\% of the commuters obtaining 0.45$\leq$PG$\leq$0.6 and 0.7$\leq$PG$\leq$0.8, respectively. When $x{=}9,500$, i.e., the maximum possible count a location could have, there is no variance in the data and MIA becomes impossible, yielding PG${=}$1. For SFC ($x \in$ \{2, 5, 10, 50, 100, 250\}), DGFR requires larger range sizes to have an effect on PG. For instance, when $x{=}50$ the PG highly varies (0.35$\leq$PG$\leq$0.8 for 50\% of the cabs) and grows slowly and more consistently as $x$ increases (mean PG 0.68 when $x{=}100$) until reaching PG${=}1$ for all users when using the maximum possible range size ($x{=}250$).

\descrit{Data Generalization with Adaptive Ranges (DGAR).} Second, aiming to increase utility, we evaluate an approach in which the range size is tailored to each location. Concretely, for {\em each location} time-series we consider the range between its minimum and maximum value over time and divide it in $x' \in$ \{1, 2, 4, 8, 16\} equal sub-ranges.
For TFL, publishing 1 or 2 sub-ranges results in a mean privacy gain of 0.91 and 0.71, respectively. When increasing $x'$, PG decreases: with $x'{=}4$ a few outliers are no longer protected from MIA. Similarly, for SFC, publishing 1 sub-range per location yields a PG between 0.55 and 0.85 for 50\% of the cabs, while with $x'{=}2$, the mean PG already drops to 0.25. This means that, as soon as information about the evolution of the location aggregates over time is revealed, cabs begin to be exposed to the attack.

\subsubsection{Hiding}\label{sec:hiding}

Another common approach used in location privacy literature is to hide (i.e., exclude) some spatio-temporal data points, by either suppressing or sampling them~\cite{HohGXA07,shokri2011quantifying}. Typically, \textit{sensitive} points are removed and the released locations are not perturbed with any kind of noise.

\descr{Suppressing Small Counts (SSC).} We first try suppressing points with small counts, i.e., assigning zeros to those location time-series points whose aggregate is under a certain threshold $k$. This approach satisfies the notion of $(k, 0)$-crowd-blending privacy, introduced by Gehrke et al. in~\cite{gehrke2012crowd}.\footnote{$k$-crowd-blending sanitization ensures that the data of each individual $u$ in a database \textit{blends} with that of $k-1$ other individuals, i.e., the output of a sanitization mechanism is \textit{indistinguishable} if $u$'s data is replaced by that of any of the other individuals.} For TFL ($k \in $ \{2, 5, 10, 20, 40, 80\}), suppressing counts with values smaller than 10 does not yield any privacy protection. As $k$ increases, we do observe some gain in privacy; e.g., PG${=}$0.38 for $k{=}20$, reaching 0.75 (resp., 0.93) when $k$ is set to 40 (resp., 80). 

The SFC dataset is much denser, thus, it requires higher $k$ to provide protection (we configure $k \in$ \{2, 5, 10, 20, 40, 80, 160\}). Surprisingly, only when $k{=}80$ the PG increases to 0.4 on average, remaining smaller than 0.2 for 25\% of the cabs. For $k{=}160$, mean PG increases to 0.78 while it is greater than 0.5 for most cabs.

\descr{Suppressing Less Popular Locations/Timeslots (SLP).} Inspired by the feature analysis (cf.~Section~\ref{sec:understanding}), which shows that some commuters/cabs are more ``distinguishable'' in the aggregates because they contribute events in less popular locations or times, we also experiment with suppressing those data points. We consider suppressing a fraction $z \in$ \{0.2, 0.4, 0.6, 0.8\} of the least popular locations and timeslots. However, this does \emph{not provide significant gains}. For instance, in TFL, suppressing 80\% of the least popular locations/timeslots still yields a PG between 0.2 and 0.3 for 50\% of the oysters. Similarly, in SFC, only some outliers are protected when we retain 20\% of the most popular locations/timeslots ($z{=}0.8$). This reinforces the conclusion that the counts of the busiest locations/times have most of the information helping MIA.

\descr{Sampling (SMP).} Another factor playing a part in the success of MIA is the amount of data users contribute to the aggregation. Therefore, we consider sampling as a means to reduce the amount of users' data. We remove a percentage $w \in $ \{0.2, 0.4, 0.6, 0.8\} of users' data points at random, and release the aggregate location time-series computed on the sampled trajectories. For TFL, this defense offers some privacy protection: the mean PG is $0.6$ when $w{=}0.4$ (with only a few outliers not so well protected), and increases up to $0.9$ when removing 80\% of the users' points. Thus, sampling might be a promising defense strategy against MIA on sparse datasets. Unsurprisingly, this approach does not work nearly as well on the denser SFC dataset. Here, PG is negligible even when 60$\%$ of the events are randomly removed, and is between 0.2 and 0.3 for 50\% of the cabs when 20\% of their points are retained.

\subsubsection{Perturbation}\label{sec:perturbation}

Next, we study the effect of perturbing the values of the aggregate location time-series, with carefully crafted \textit{noise}. The state-of-the-art method for releasing aggregate statistics free from inferences is to satisfy differential privacy (DP)~\cite{dwork2008differential}. In this setting, Acs and Castelluccia~\cite{acs2014case} present an algorithm tuned to the density of Paris for releasing aggregate statistics from a telecommunication service provider's dataset, while Quercia et al.~\cite{quercia2011spotme} use randomized response to let an untrusted aggregator privately learn the number of people in geographic locations. However, previous work~\cite{pyrgelis2018knock,pyrgelis2017does,to2016differentially} has shown that DP techniques ultimately yield poor utility on high-dimensional settings. To mitigate this problem, the noise addition can be configured to achieve weaker privacy guarantees, such as crowd-blending privacy~\cite{gehrke2012crowd}, while retaining better utility levels. For instance, To et al.~\cite{to2016differentially} apply this notion to release privacy-preserving location entropy statistics.

\descr{Perturbing Small Counts (PSC).} We first add noise sampled from the Laplace distribution with scale $O(1 / \epsilon')$ \emph{only} to small counts, i.e., counts of the aggregate location time-series that are smaller than a threshold $k$. 
This achieves $(k, \epsilon')$-crowd-blending privacy~\cite{gehrke2012crowd}. 
We range $k$ as for SSC and configure $\epsilon'$ to 1.0 for TFL and to 0.1 for SFC, since we expect that the denser SFC dataset requires larger scale noise addition to obtain some privacy protection. For TFL, this approach results in reasonable privacy gain. For instance, with $k{=}5$ the mean PG is 0.55 (with some outliers having less protection), while with $k{=}40$ the PG is higher than 0.6 for all commuters. On the contrary, for SFC, this mechanism does not offer much protection unless $k{=}160$, where 0.6$\leq$PG$\leq$0.7 for 50\% of the cabs. %

\descr{Fourier Perturbation Algorithm (FPA).} We then experiment with FPA~\cite{rastogi2010differentially}, a differentially private mechanism tailored to time-series settings. %
FPA operates as follows: a time-series is first transformed to the frequency domain using the Discrete Fourier Transform (DFT), and $l$ Fourier coefficients $F_l$ are retained ($l$ is an algorithm parameter). Then, $F_l$ is perturbed with noise sampled from the Laplace distribution, with scale $O(\sqrt{l}\cdot \Delta f_{2} / \epsilon)$ and padded with zeros to the size of the original time-series (note that $\Delta f_{2}$ depicts the $\ell_2$ norm of the users' sensitivity). Finally, the inverse DFT is performed on $F_l$ to obtain the perturbed time-series. As discussed in~\cite{rastogi2010differentially}, FPA provides $\epsilon$-DP guarantees.

We set $\epsilon$ in \{0.01, 0.1, 1.0, 10.0\}. As expected, PG is higher for smaller values of $\epsilon$ (i.e., stronger DP privacy guarantees). For TFL, with $\epsilon{\leq}0.1$ PG reaches 1.0, while with $\epsilon{=}1.0$ it is above 0.85 for all the users. For larger values of $\epsilon$, (i.e., $10.0$), PG drops between 0.4 and 0.5 for 50\% of the commuters. Similarly, for SFC, the mean PG is very high for $\epsilon$ values up to $1.0$ (e.g., PG$\geq$0.75 for all the cabs when $\epsilon{=}1.0$) but only a few cabs are well protected when $\epsilon{=}10.0$.

\subsubsection{Combining Hiding and Perturbation}\label{sec:combination} 

Finally, we investigate whether combining defense strategies can improve the privacy gain. In particular, we focus on the combination of sampling with perturbation which has been suggested in previous work~\cite{gehrke2012crowd,Li:2012:SAD:2414456.2414474}. For instance, Gehrke et al.~\cite{gehrke2012crowd} show that introducing a random sampling step before the application of a crowd-blending privacy mechanism (e.g., PSC) increases the adversary's uncertainty and achieves a stronger privacy notion, namely, zero-knowledge privacy~\cite{gehrke2011towards}.
Similarly, Li et al.~\cite{Li:2012:SAD:2414456.2414474} suggest that a random sampling step can amplify the privacy offered by a differentially private mechanism, thus, we also combine sampling with FPA. The rationale is that random sampling decreases the amount of data contributed by users, thus, it reduces the sensitivity of the aggregation function and less noise is required to obtain the differential privacy guarantees.

\descr{Sampling \& Perturbing Small Counts (SPSC).} For TFL, we set $k$ to 5 and $\epsilon'$ to 1.0, while for SFC, we set $k$ to 20 and $\epsilon'$ to 0.1. For both datasets, we range the sampling parameter $w $ in \{0.2, 0.4, 0.6, 0.8\}. We find that the introduction of a random sampling step boosts the PG of the perturbation mechanism. For TFL, when we retain 80$\%$ of the data (i.e., $w{=}0.2$), the PG is higher than 0.68 for all the users and 0.85 on average. This is a 4$\times$ and 2$\times$ increase compared to SMP or PSC alone. Furthermore, with $w{=}0.6$ or $0.8$, PG almost reaches 1. On the SFC dataset, we observe once again that more data needs to be removed to obtain some PG for the cabs. In particular, sampling boosts mean privacy gain to 0.21 when $w{=}0.6$, a 4- and 21-fold increase compared to SMP and PSC alone. When $w{=}0.8$, PG is between 0.45 and 0.6 for 50$\%$ of the cabs.

\descr{Sampling \& Fourier Perturbation Algorithm (SFPA).} For both datasets, we configure FPA's privacy budget $\epsilon$ to the least conservative setting, i.e., $\epsilon{=}10.0$, while varying $w$, to observe its amplifying effect on PG. For TFL, when we retain 80$\%$ of the data (i.e., $w{=}0.2$) the mean PG is 0.68, which is approx.~2.5 times or 1.5 times higher than SMP or FPA, respectively, alone. As we increase $w$, the PG is increasing further, e.g., with $w{=}0.8$, PG $\geq$ 0.8 for all commuters. For SFC, higher sampling rates are needed to get similar privacy levels. When $w{=}0.4$, a few outliers are not protected from MIA while when $w{=}0.8$, 0.7$\leq$PG$\leq$0.8 for 50\% of the cabs.

\subsection{Unsuitable Defenses}
In theory, one could also attempt to add {\em dummy} locations to obfuscate users' trajectories~\cite{Meyerowitz09}. However, it is well-known that generating plausible dummy locations is extremely hard, as those can be easily filtered by the adversary by exploiting statistical correlations with real locations, and thus ultimately provide no protection~\cite{ChowG09}. 

Another approach would be to generate privacy-preserving {\em synthetic traces}~\cite{bindschaedler2016synthesizing,machanavajjhala2008privacy,he2015dpt,gursoy2018utility,mir2013dp} 
and compute the aggregate statistics on them rather than on the actual locations of the users. However, when attempting to implement and evaluate these methods, we found that the synthetic trajectories generated using techniques presented in~\cite{he2015dpt,gursoy2018utility} do not preserve the time dimension, while those proposed in~\cite{machanavajjhala2008privacy,mir2013dp} only work for computing origin-destination commute distances.

Bindschaedler and Shokri~\cite{bindschaedler2016synthesizing} also support {\em plausible deniability} with respect to whether a real trace (a ``seed'') was part of the training set used by a privacy-preserving generative model to produce synthetic traces. Given a set of synthetic traces, an adversary cannot learn which locations the seed contributors have visited or whether a user with certain semantic habits is in the seed dataset. 
This is a different goal than the one we consider, i.e., given aggregate location statistics, preventing an adversary from inferring whether a user contributed to them.
Moreover, when the goal is to release traces~\cite{bindschaedler2016synthesizing}, one can afford to aggressively sample points from the trajectories in order to make the system scale. 
In the cases considered in this paper, where the goal is to obtain fine-grained aggregate statistics we cannot remove as many intermediate points.
Not using aggressive sampling remarkably increases the computational cost of the generation process and ultimately makes it impossible to properly evaluate this strategy.

\subsection{Take Aways}
Our analysis of the defenses shows that {\em spatio-temporal generalization}, a technique commonly used to protect privacy for mobility trajectories~\cite{krumm2007inference,GruteserG03,de2013unique}, does not yield meaningful protection against MIA on aggregate location time-series: both space and time dimensions contain information that is useful for the attack. On the contrary, {\em data generalization} approaches like discretizing the counts of the time-series do provide acceptable privacy levels. %

Second, we find that {\em hiding} techniques (e.g., suppression or sampling) yield reasonably high privacy levels when the input signal is sparse (as for TFL), although they do not work as well on dense datasets (SFC). 
Regardless, suppressing locations/timeslots based on popularity does not improve privacy, as busy ROIs/times are significantly informative for the attack. 

Finally, {\em perturbation} techniques configured to guarantee differential privacy achieve, as expected, very high gains in privacy. However, similar levels of protection can actually be reached with less noise using techniques like crowd-blending sanitization. Also, combining both sampling and perturbation can significantly amplify the privacy gain since the former increases the adversary's uncertainty while reducing the sensitivity of the aggregation function.

\section{Privacy--Utility Trade-Offs}\label{sec:priv-util-tradeoff}
Our last set of measurements studies the impact of the evaluated defense strategies on {\em utility}. 
While privacy is quantified via the ability of mitigating MIA (i.e., the Privacy Gain), we measure utility vis-\`a-vis analytics tasks run on aggregate location time-series:
\begin{enumerate}
\item Forecasting traffic volumes in Regions of Interest~\cite{jiang2018deep}, %
\item Mining interesting locations/discovering hotspots~\cite{karamshuk2013geo}, %
\item Inferring maps or labeling locations~\cite{liu2012mining}, and 
\item Detecting mobility anomalies~\cite{pyrgelis2016privacy}. 
\end{enumerate}

Specifically, we consider utility metrics that capture the key characteristics of the aggregate data enabling that application.

To ease presentation, in Figures~\ref{fig:mre-utility}--\ref{fig:correlation-utility}, we plot the privacy gain vs.~the utility loss (i.e., the decrease in utility compared to performing the same task on \textit{raw} aggregate location time-series).\footnote{In Appendix~\ref{app:utility-tables}, we include several tables indicating how the configuration of each defense affects the utility metrics under consideration.} Ideally, we would like to obtain data points on the upper left corner of the plots, i.e., where the privacy gain is high and the utility loss is low.
In the rest of the section, for each analytics task considered, we describe the task and we discuss the suitability of the defenses.

\subsection{Traffic Forecasting: Aggregates Error}
Aggregate location time-series are often used for forecasting traffic in ROIs~\cite{jiang2018deep,hoang2016forecasting}. We measure the utility loss by quantifying the effect of a defense on the {\em precision} of the data release, i.e., the error the defense introduces, as that would inevitably effect the forecast as well. We use the Mean Relative Error (MRE) over the whole time-series or a percentage thereof (e.g., the MRE over the 10\% busiest ROIs). Given two time-series $Y$ and $Y'$, of length $|\mathcal{T}|$, denoting the aggregates before and after a defense has been applied, we calculate: \vspace*{-0.15cm}
\begin{equation}\label{eq:mre}
\small
\text{MRE}(Y, Y') = \frac{1}{|\mathcal{T}|} \sum_{t \in \mathcal{T}} \frac{ | Y'_{t} - Y_{t} | }{ \text{max}(\gamma, Y_{t}) }\vspace*{-0.125cm}
\end{equation}
($\gamma$ is a sanity bound mitigating the effect of very small counts).

\begin{figure*}[t]
\centering
\begin{subfigure}{0.3\textwidth}
\includegraphics[width=1.0\textwidth]{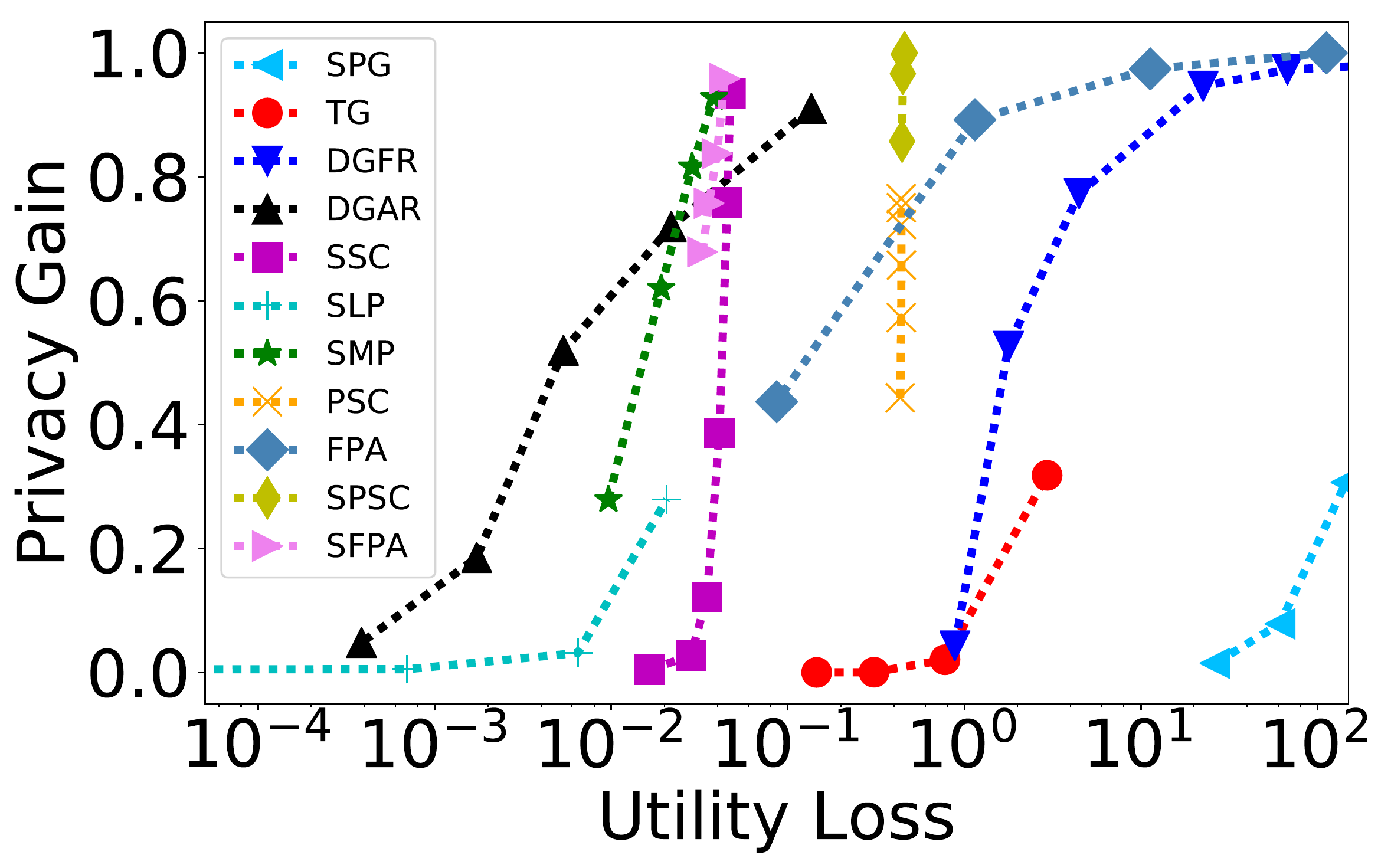}
\caption{TFL - MRE}
\label{fig:tfl-mre}
\end{subfigure}
\begin{subfigure}{0.3\textwidth}
\includegraphics[width=1.0\textwidth]{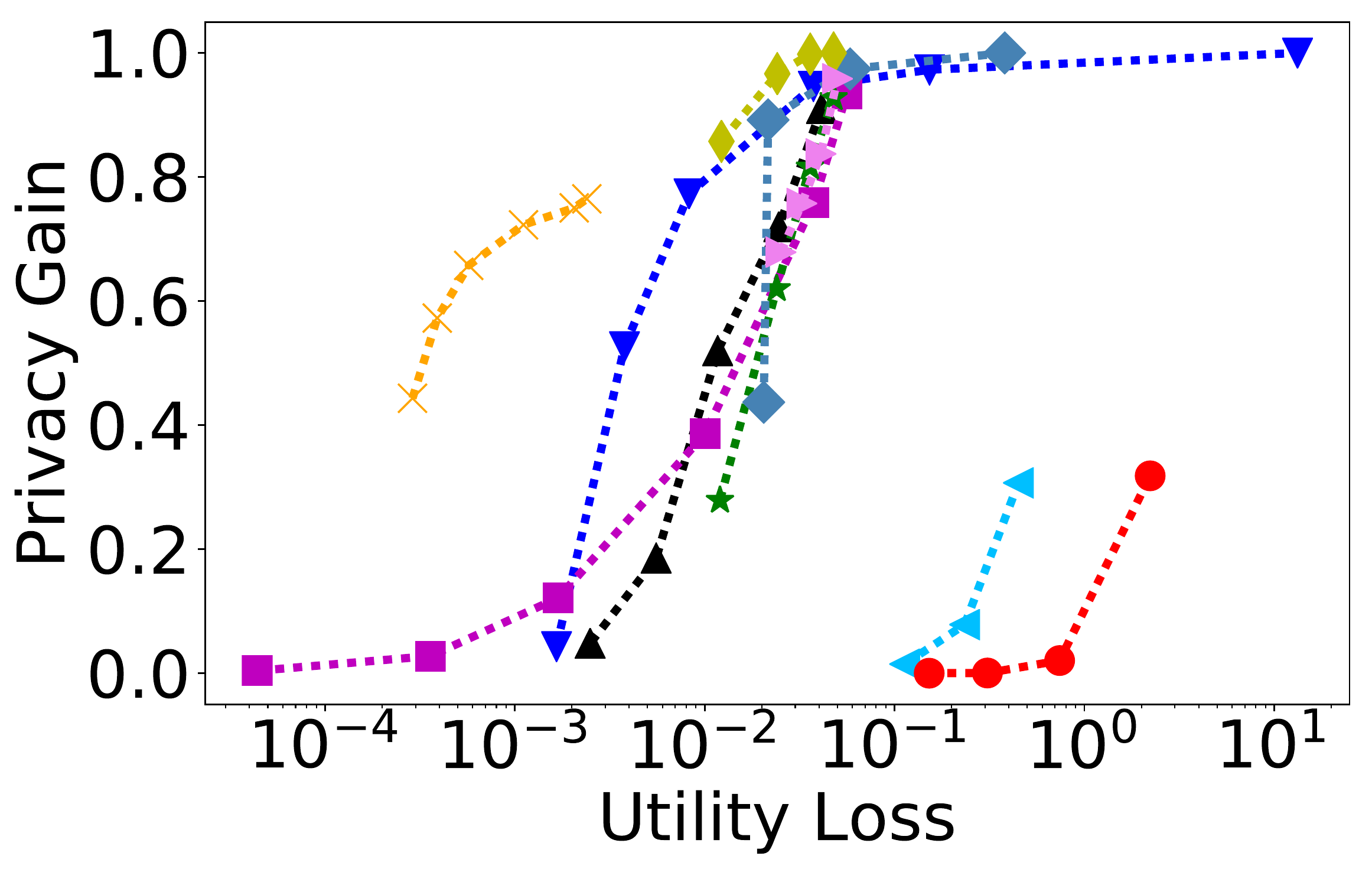}
\caption{TFL - MRE 10$\%$}
\label{fig:tfl-mre10}
\end{subfigure}
\\
\begin{subfigure}{0.3\textwidth}
\includegraphics[width=1.0\textwidth]{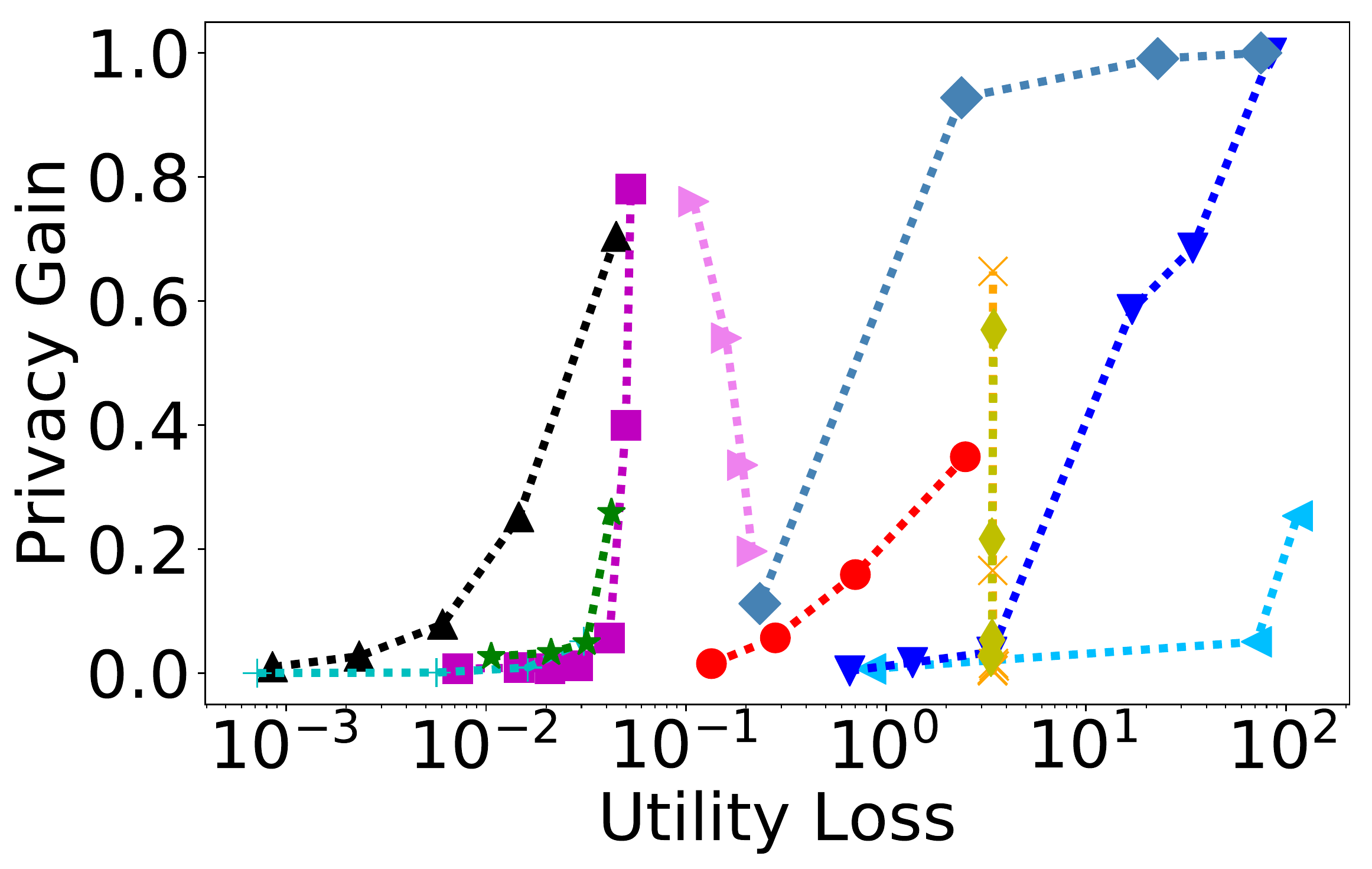}
\caption{SFC - MRE}
\label{fig:sfc-mre}
\end{subfigure}
\begin{subfigure}{0.3\textwidth}
\includegraphics[width=1.0\textwidth]{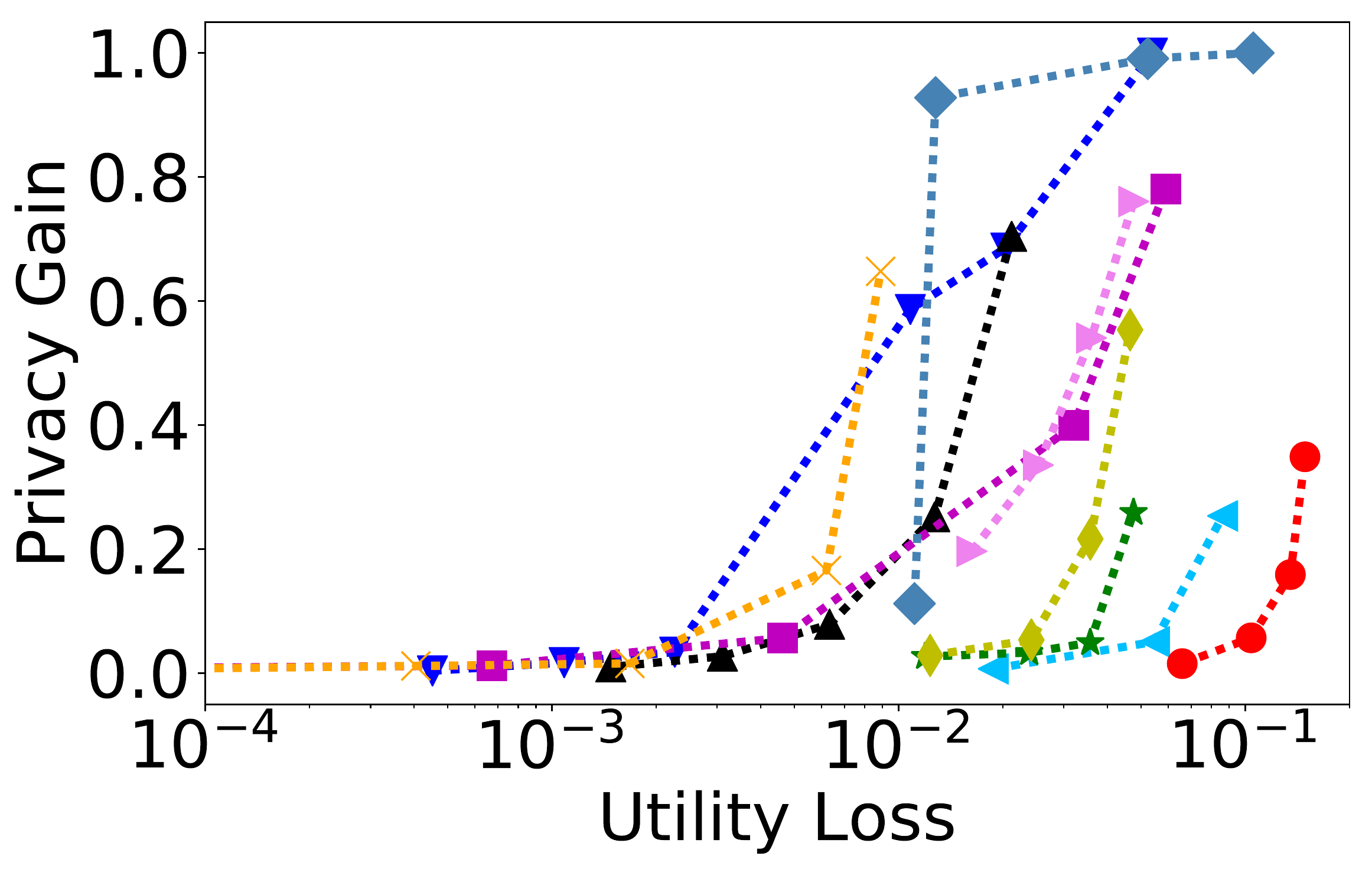}
\caption{SFC - MRE 10$\%$}
\label{fig:sfc-mre10}
\end{subfigure}
\caption{Privacy--Utility Trade-off: Traffic Forecasting (Aggregates Error).}
\label{fig:mre-utility}
\end{figure*}

Figure~\ref{fig:mre-utility} plots the results for this task. For TFL, defenses like data generalization with adaptive ranges (DGAR), sampling alone (SMP) or combined with FPA perturbation (SFPA), and small count suppression (SSC), yield reasonable privacy--utility trade-offs (Figure~\ref{fig:tfl-mre}). If one is interested in performing predictive analytics only on the busiest stations (e.g., MRE 10$\%$ -- Figure~\ref{fig:tfl-mre10}), defenses such as perturbing small counts (PSC) or data generalization with fixed ranges (DGFR) yield better trade-offs, with the former performing better than the latter in terms of utility (but worse for privacy). In this setting, if slightly higher utility loss can be tolerated, sampling and perturbation of small counts (SPSC) as well as FPA can provide better privacy (i.e., PG between 0.8 and 1.0). 

For SFC, DGAR or SSC yield small MRE overall (Figure~\ref{fig:sfc-mre}) and could be used well for predictive analytics. Nonetheless, to forecast traffic in the top 10$\%$ of SFC regions (Figure~\ref{fig:sfc-mre10}), FPA or PSC yield an efficient privacy--utility trade-off balance (0.9 PG and $10^{-2}$ utility loss for the former and 0.64 PG and $9\cdot10^{-3}$ utility loss for the latter).

\subsection{Hotspot Discovery: Prediction Accuracy \& Rank Correlation}

Analysts are often interested in predicting \textit{location hotspots} over time~\cite{zheng2009mining}; this is particularly useful for journey planning and/or resource allocation.
For instance, authorities need to learn which stations are the busiest in certain hours of a day to allocate staff accordingly, or to suggest alternative routes to commuters. 
Hotspot discovery is also crucial to identify the optimal location and time to place advertisements, open new shops, etc.~\cite{smartsteps,karamshuk2013geo}. For this task, we measure utility as follows: we use the aggregates after applying a defense to predict the busiest $10\%$ ROIs at each timeslot of the inference week, and calculate the F1 score of the predictions: \vspace*{-0.1cm}
\begin{equation}\label{eq:f1}
\small
\text{F1} = 2 \cdot\frac{\text{PPV} \cdot \text{TPR}}{\text{PPV} + \text{TPR}}
\vspace*{-0.1cm}
\end{equation}
where $\text{PPV} = \frac{\text{TP}}{\text{TP} + \text{FP}}$ is the precision and $\text{TPR} = \frac{\text{TP}}{\text{TP} + \text{FN}}$ is the recall of the predictions, with $\text{TP}, \text{FP}$ and $\text{FN}$ indicating the true/false positives and false negatives, respectively. 

F1 score quantifies how successful hotspot prediction is in each timeslot, but does not capture if the ordering of the hotspots is preserved. This might be important for resource planning, e.g., a taxi company assigning vehicles to locations sorted by client demand. Thus, we also calculate the Kendall rank correlation coefficient, a measure of correspondence between two rankings, whereby values closer to $1$ indicate strong agreement and those closer to $-1$ strong disagreement. More precisely, given two rankings, $X$ and $X'$, the Kendall rank correlation $\tau(X,X')$ is computed as:
\begin{equation}\label{eq:kendall}
\tau(X,X') = \frac{P - Q}{ \sqrt{ (P + Q + T) \cdot (P + Q + U)} }
\end{equation}
where $P$ is the number of concordant pairs, $Q$ that of discordant pairs, $T$ the number of ties only in $X$, and $U$ the number of ties only in $X'$. If a tie occurs for the same pair in both $X$ and $X'$, it is not added to either $T$ or $U$~\cite{kendall1945treatment}.

\begin{figure*}[t]
\centering
\begin{subfigure}{0.3\textwidth}
\includegraphics[width=1.0\textwidth]{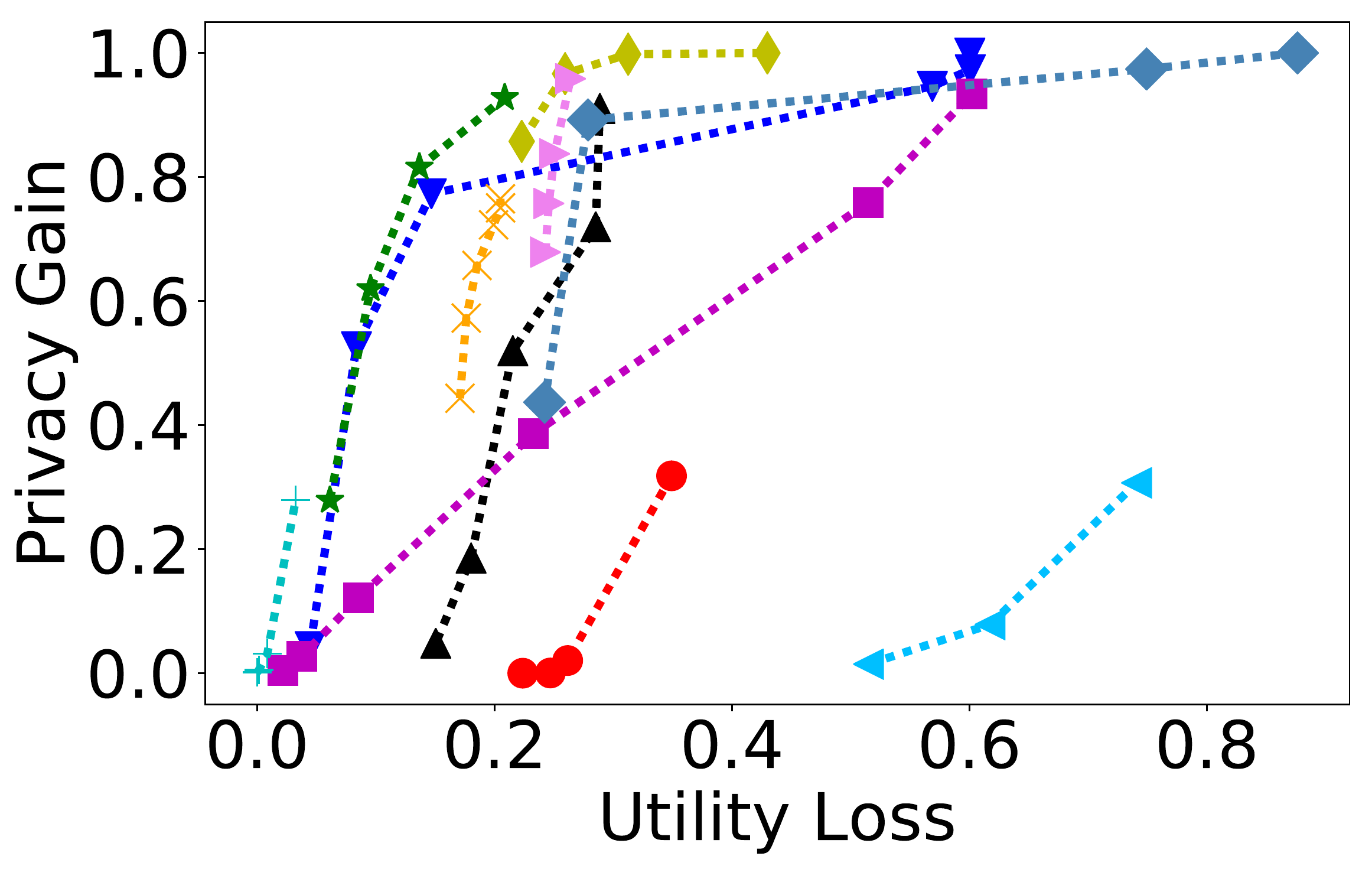}
\caption{TFL - F1}
\label{fig:tfl-f1}
\end{subfigure}
\begin{subfigure}{0.3\textwidth}
\includegraphics[width=1.0\textwidth]{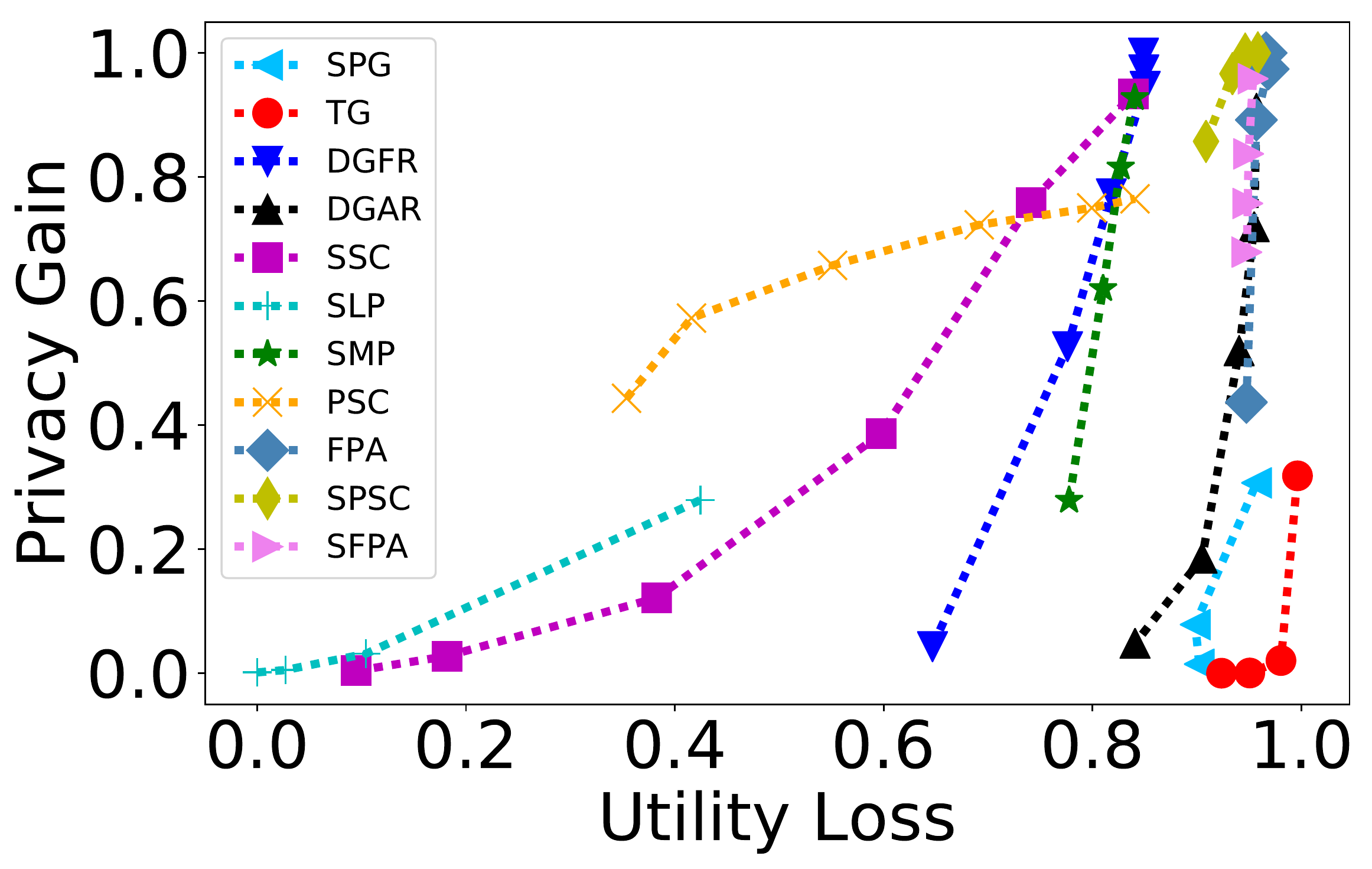}
\caption{TFL - $\tau$}
\label{fig:tfl-tau10}
\end{subfigure}
\\
\begin{subfigure}{0.3\textwidth}
\includegraphics[width=1.0\textwidth]{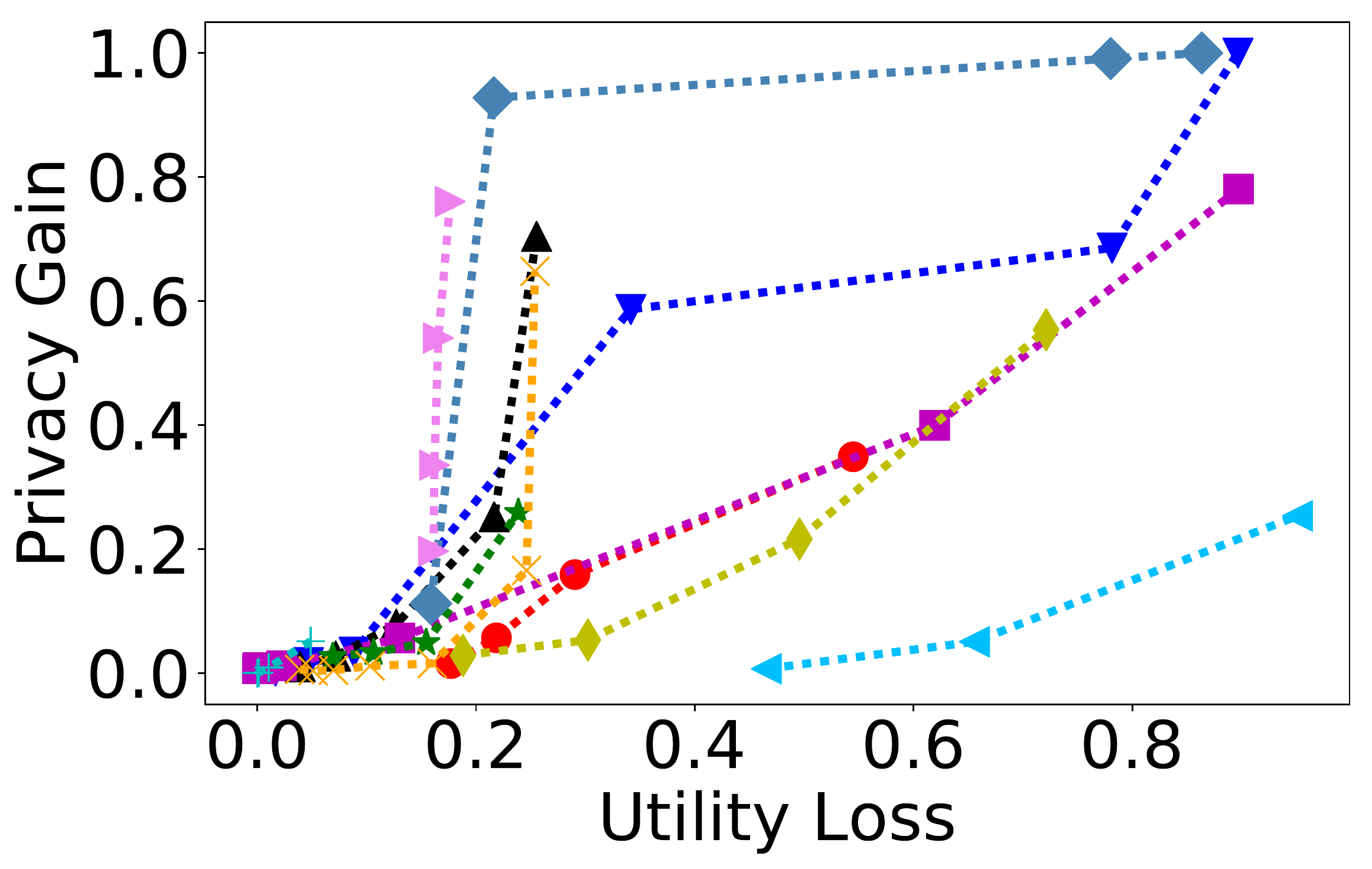}
\caption{SFC - F1}
\label{fig:sfc-f1}
\end{subfigure}
\begin{subfigure}{0.3\textwidth}
\includegraphics[width=1.0\textwidth]{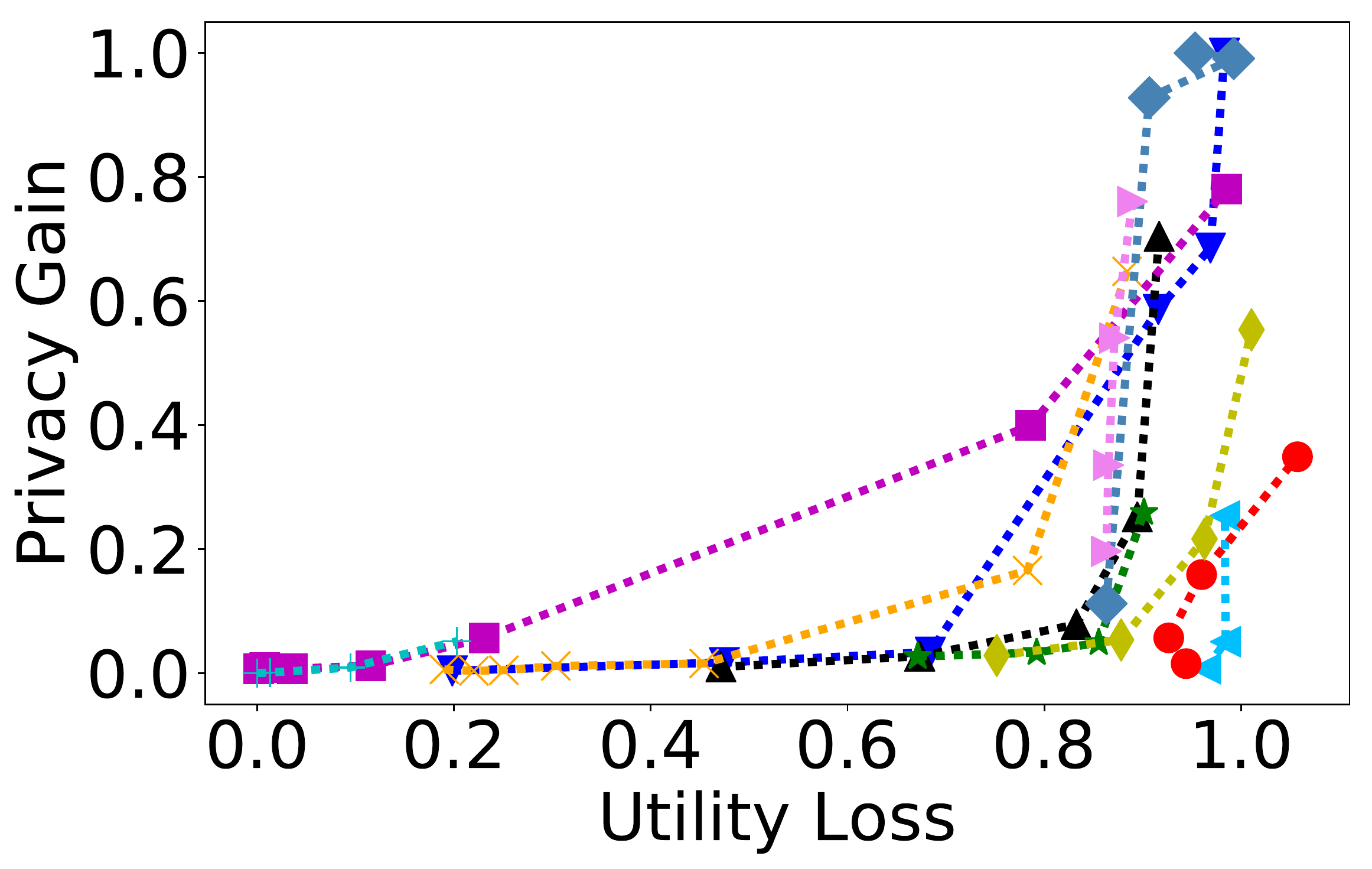}
\caption{SFC - $\tau$}
\label{fig:sfc-tau10}
\end{subfigure}
\caption{Privacy--Utility Trade-off: Hotspot Discovery (Prediction Accuracy \& Rank Correlation).}
\label{fig:prediction-utility}
\end{figure*}

In the case of TFL, Figure~\ref{fig:tfl-f1} shows that DGFR and SMP have small utility loss (0.1--0.15) and relatively high privacy gain (0.5--0.8), indicating that they are indeed suitable for hotspot discovery tasks. However, if the ranking of the top stations is crucial, then, defense strategies such as SSC or PSC outperform the others (Figure~\ref{fig:tfl-tau10}). In SFC, for hotspot prediction, FPA and SFPA perform best as they yield higher PG for similar levels of utility. The same observation holds for ranking the top locations (Figure~\ref{fig:sfc-tau10}), even though SSC could be used if one is willing to sacrifice some privacy for slightly better utility (0.4 PG and 0.75 utility loss).

\subsection{Map Inference: Distribution Similarity} 

Tasks like map inference -- i.e., inferring road maps from GPS traces~\cite{liu2012mining} or labeling locations~\cite{ye2011semantic} -- rely on the fact that certain locations are more frequently visited than others~\cite{bindschaedler2016synthesizing}. Thus, to evaluate utility in this setting, we use the Jensen-Shannon (JS) divergence, which estimates the similarity between two probability distributions.
This captures whether the distribution of location visits is preserved (for each timeslot) after applying a defense. JS is a smoothed version of the Kullback-Leibler (KL) divergence that is symmetric and always defined. Given two probability distributions, $V$ and $W$, the JS-divergence is calculated as:%
\begin{equation}\label{eq:js}
\small
\text{JS}(V||W) = \frac{1}{2} \cdot \text{KL}(V||Z) + \frac{1}{2} \cdot \text{KL}(W||Z) %
\end{equation}
where $Z = \frac{1}{2} \cdot (V + W)$. JS is a value between 0 and 1 with larger values indicating bigger distance between the distributions (i.e., worse utility for map inference).

\begin{figure}[t]
\centering
\begin{subfigure}{0.3\textwidth}
\includegraphics[width=1.0\textwidth]{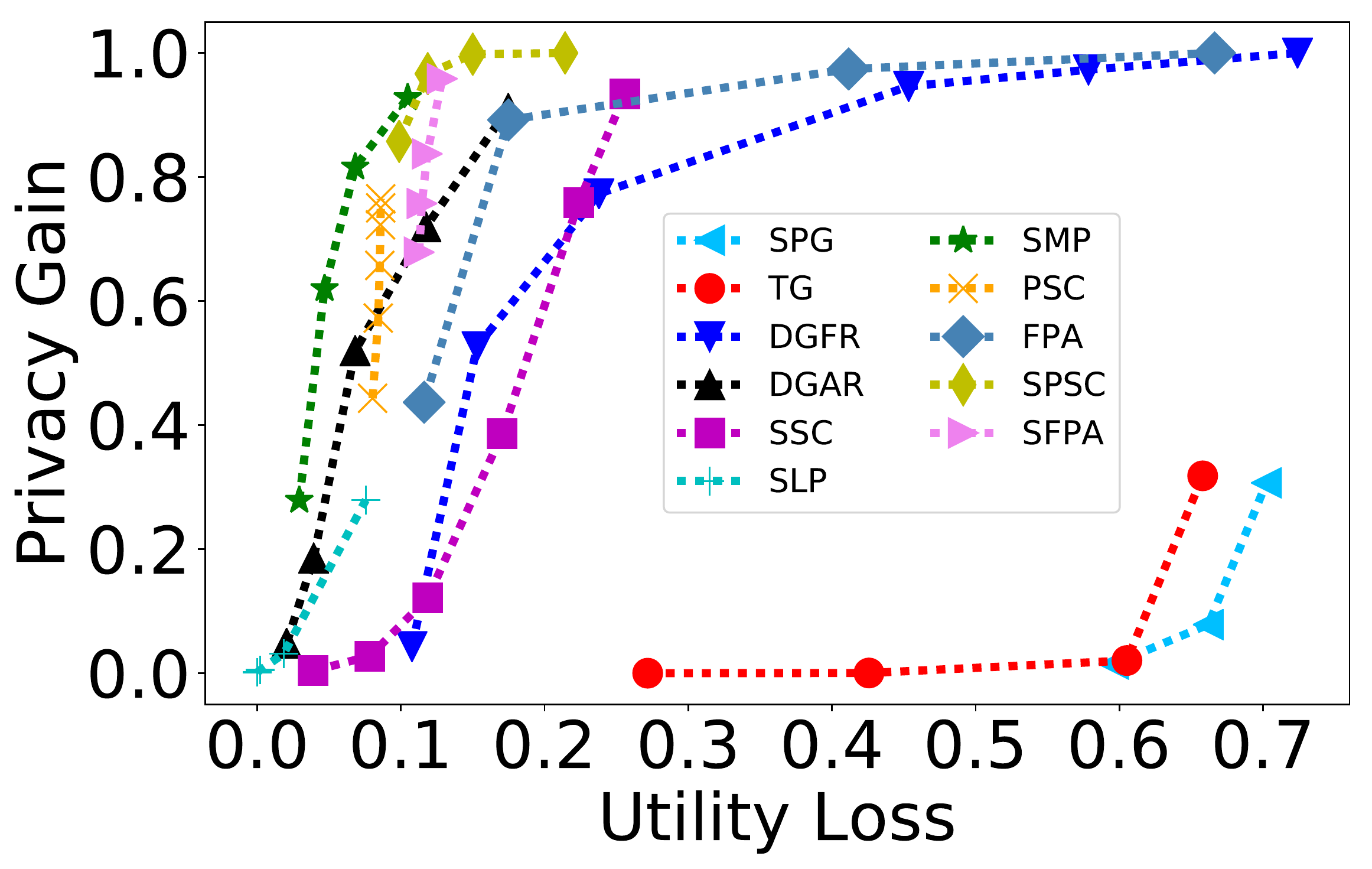}
\caption{TLF - JS}
\label{fig:tfl-js}
\end{subfigure}
\hspace{0.2cm}
\begin{subfigure}{0.3\textwidth}
\includegraphics[width=1.0\textwidth]{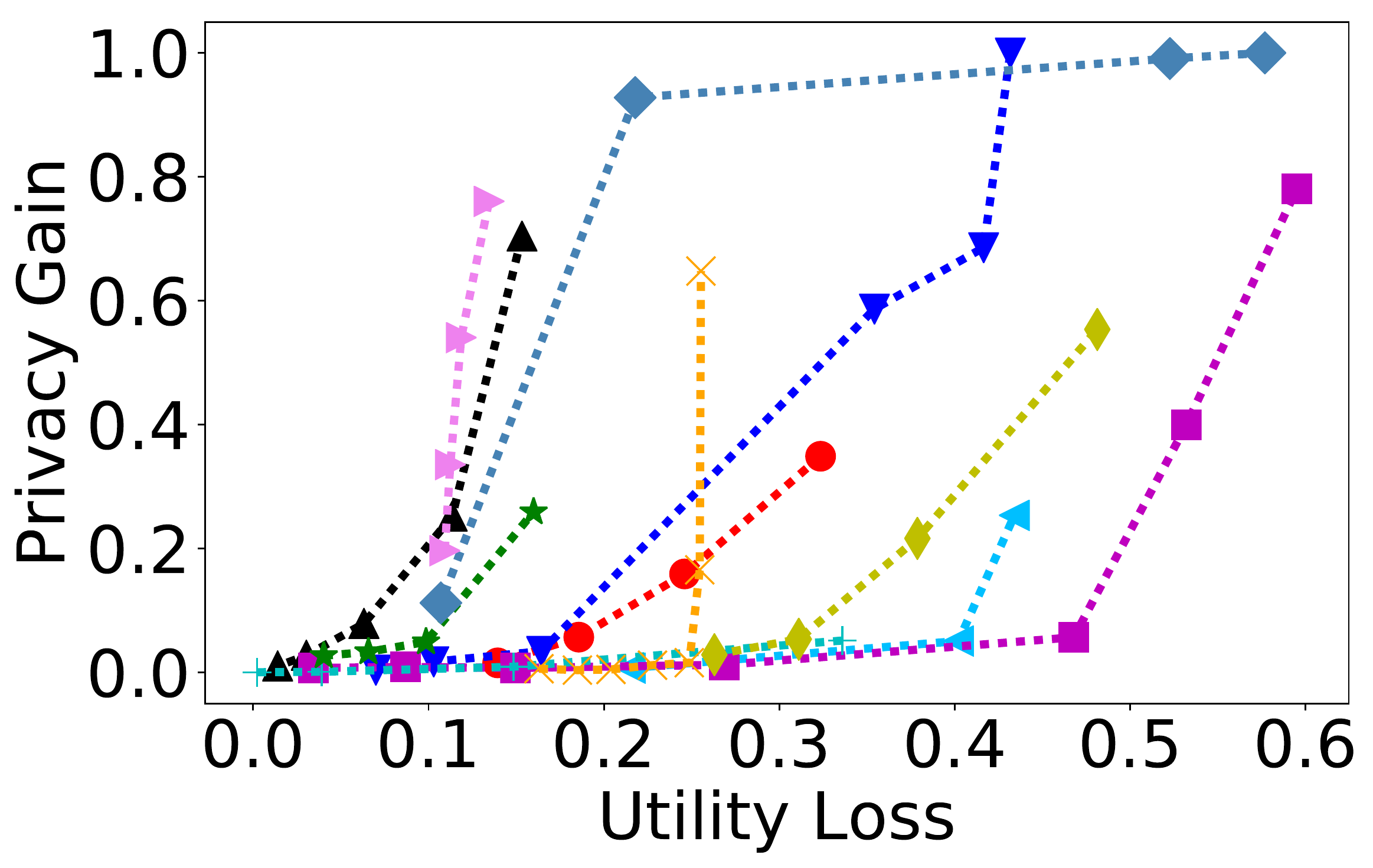}
\caption{SFC - JS}
\label{fig:sfc-js}
\end{subfigure}
\caption{Privacy--Utility Trade-off: Map Inference (Distribution Similarity).}
\vspace*{-0.2cm}
\label{fig:distribution-utility}
\end{figure}

Figure~\ref{fig:distribution-utility} visualizes the privacy--utility trade-off of the various defenses for map inference tasks, for both datasets. For TFL, Figure~\ref{fig:tfl-js} shows that a few defenses, including sampling without or with FPA (SMP or SFPA), DGAR, PSC, and SPSC do achieve good trade-offs. For instance, sampling yields a 0.06 utility loss for a mean privacy gain of 0.8, while, when combined with small count perturbation, PG reaches 0.96 for a utility loss of 0.11. For SFC, DGAR as well as SFPA yield privacy gains between 0.25 and 0.75 with approx.~0.1 utility loss (Figure~\ref{fig:sfc-js}). Higher privacy levels achieved by FPA or DGFR only come with increase in utility loss, while other defenses such as sampling with small count perturbation or small count suppression (SPSC or SSC) yield worse utility without actually improving privacy.

\subsection{Anomaly Detection: Correlation}

\begin{figure}[t]
\centering
\begin{subfigure}{0.3\textwidth}
\includegraphics[width=1.0\textwidth]{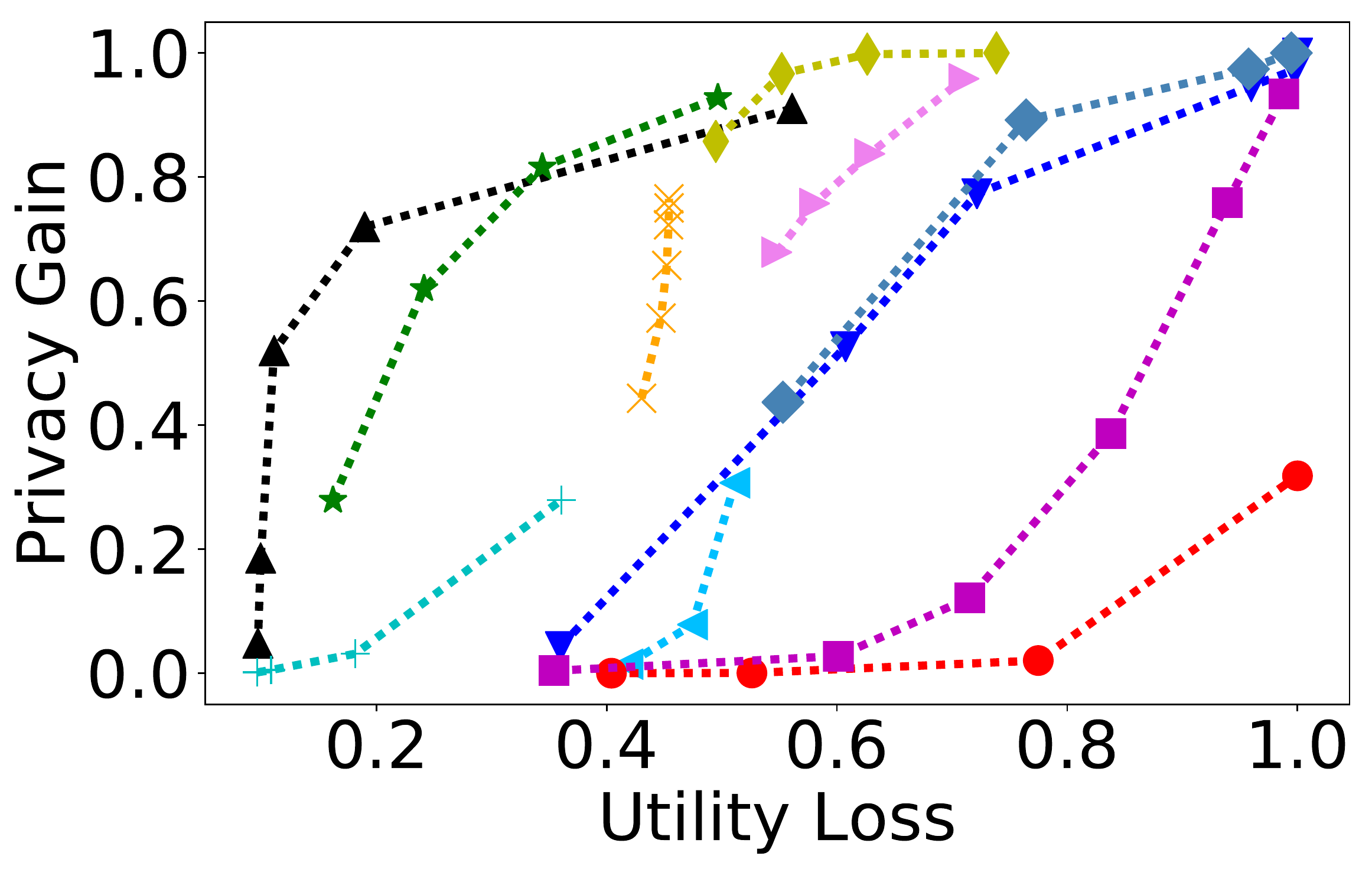}
\caption{TFL - $r$}
\label{fig:tfl-r}
\end{subfigure}
\hspace{0.2cm}
\begin{subfigure}{0.3\textwidth}
\includegraphics[width=1.0\textwidth]{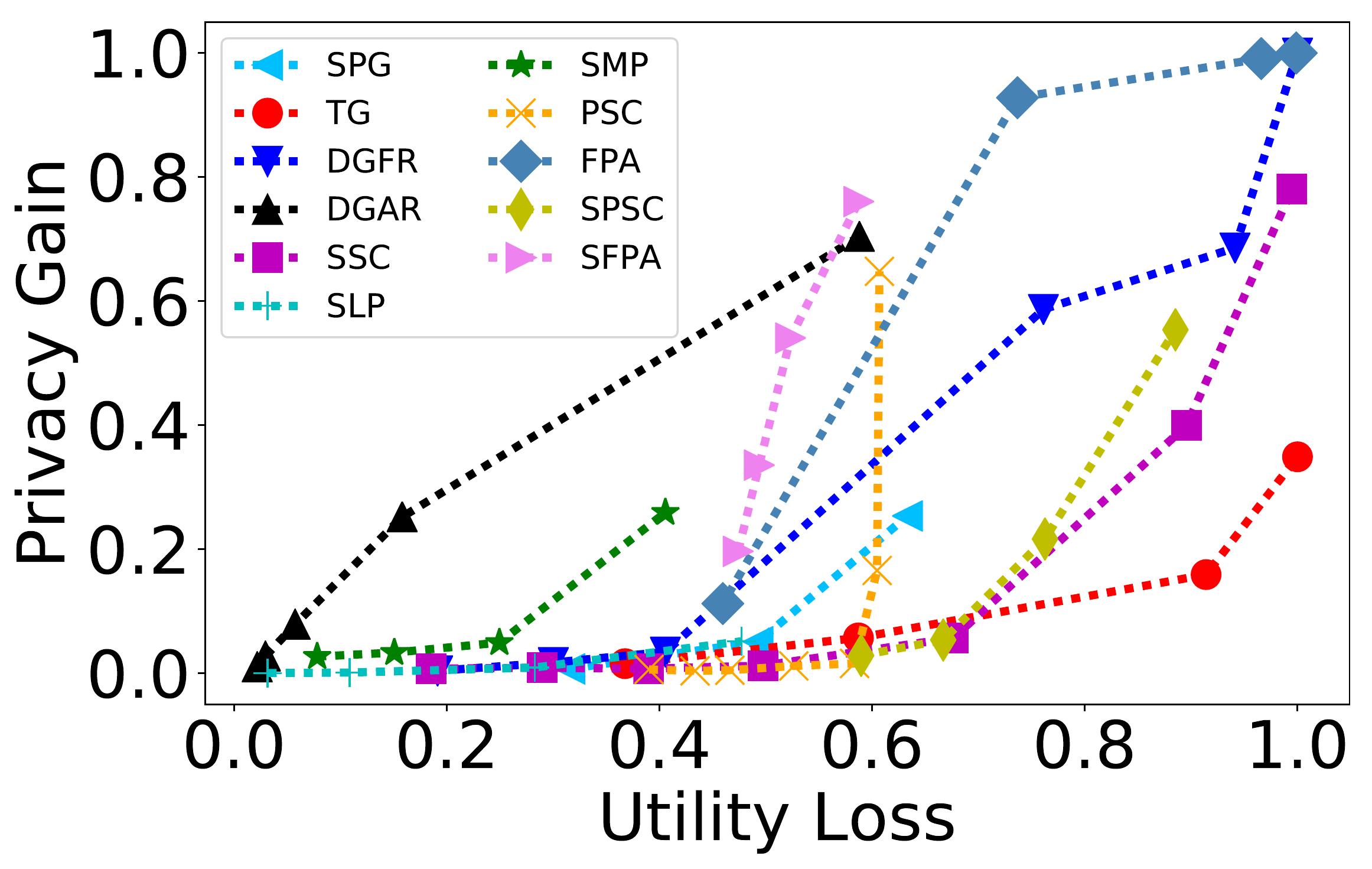}
\caption{SFC - $r$}
\label{fig:sfc-r}
\end{subfigure}
\caption{Privacy--Utility Trade-off: Anomaly Detection (Correlation).}
\label{fig:correlation-utility}
\end{figure}

Finally, analytics aiming to detect mobility anomalies~\cite{pan2013crowd} and/or improve traffic forecasting in the presence of an anomaly, require that a linear relationship between two time-series---before and after applying a defense---is preserved~\cite{pyrgelis2016privacy}. Thus, here we calculate the Pearson correlation coefficient between the perturbed and raw aggregate time-series to measure utility. The Pearson correlation varies between $-1$ and $+1$, with values closer to $1$ indicating positive linear correlation, and to $-1$ total negative correlation (values close to $0$ imply no linear correlation). Given two signals $Y$ and $Y'$, the Pearson correlation is computed as:%
\begin{equation}\label{eq:pearson}
r(Y, Y') = \frac{ \sum (Y - \mu_{Y}) \cdot ( Y' - \mu_{Y'}) }{ \sqrt{ (Y - \mu_{Y})^{2} \cdot ( Y' - \mu_{Y'} )^{2} } } %
\end{equation}
where $\mu_{X}$ is the mean of a signal $X$.

In the TFL setting, data generalization with adaptive ranges (DGAR) and sampling (SMP) offer a reasonable balance in the trade-off (Figure~\ref{fig:tfl-r}), while other defenses, such as FPA, data generalization with fixed ranges (DGFR), and small count suppression (SSC), increase privacy only if breaking the correlations. For SFC (Figure~\ref{fig:sfc-r}), FPA with or without sampling yields, resp., 0.75 PG with 0.6 utility loss and 0.92 PG with 0.7 utility loss. In this case, DGAR achieves smaller utility loss (0.18) with some decrease in privacy gain (0.25).

\subsection{Take Aways}

Our measurements highlight the high variance in trade-offs between utility and privacy for the different defenses and analytics tasks. Thus, it seems that there does not exist a unique generic defense that can preserve the utility of the analytics for arbitrary applications.

Spatio-temporal generalization yields poor utility overall, and, as discussed in Section~\ref{sec:defenses}, it anyway does not protect privacy. Other defenses, e.g., suppressing locations/timeslots based on their popularity, generally yield good utility for the analytics under consideration, but they provide poor levels of privacy. Defense strategies like data generalization, small counts suppression, sampling, perturbation, or combinations of the last two, can be configured to obtain reasonable privacy levels while still providing reasonable utility for specific applications. In particular, data generalization techniques enable analysts to perform forecasting traffic tasks, hotspot discovery, and map inference, while small count suppression can be useful towards ranking hotspots. Sampling can retain utility for tasks such as location labeling and anomaly detection, although, from a privacy perspective, it works better when the dataset is sparse.

Perturbation techniques configured to achieve strong differential privacy achieve reasonable accuracy for applications such as discovering hotspots and forecasting their traffic, while additional tasks, e.g., forecasting the traffic of less busy ROIs or detecting anomalies, are more efficient when the injected noise is tailored to achieve weaker privacy notions (e.g., crowd-blending privacy). Interestingly, our results show that combining defenses, e.g., sampling and perturbation, not only helps privacy, but also retains utility for tasks such as ranking hotspots, map inference, and anomaly detection.

\section{Related Work}\label{sec:related}

\descr{Location Privacy.} Golle and Partridge~\cite{golle2009anonymity} demonstrate the feasibility of re-identifying users by leveraging the uniqueness of their home/work places.
Shokri et al.~\cite{shokri2010unraveling} show that k-anonymity in the context of location traces is mostly ineffective, while 
Zang and Bolot~\cite{zang2011anonymization}  that anonymization of location data is, in general, extremely difficult. Furthermore, other works show that location data can be de-anonymized using data originating from network providers~\cite{wang2017fingerprint,8622184}, social-network graphs~\cite{srivatsa2012deanonymizing,ji2016general}, or check-in/review services~\cite{wang2018anonymization}.

De Montjoye et al.~\cite{de2013unique} measure the uniqueness of human mobility %
in a Call Detail Records (CDR) dataset, finding that four spatio-temporal points are enough to uniquely identify 95\% of the individuals, while Rossi et al.~\cite{rossi2015spatio} study how mobility features like speed, direction, and distance can be used to link  trajectories to specific users. Although some of these efforts also focus on understanding which characteristics of location data threaten user privacy, they do so for {\em single users'} location traces---a different setting than aggregate location time-series. Furthermore, our study investigates how defense strategies based on generalization~\cite{GruteserG03,cai2014systematic,venkatadri2018privacy}, hiding~\cite{HohGXA07,shokri2011quantifying}, and perturbation~\cite{to2016differentially} that are commonly used in location privacy literature, protect aggregate location data.

\descr{Aggregate Location Privacy.} Aggregation is often not an effective way to preserve the privacy of location data, as aggregates leak information about individual users.
Xu et al.~\cite{xu2017trajectory} reconstruct victims' location trajectories from aggregate mobility data, without any prior knowledge,
while~\cite{pyrgelis2017does} shows that aggregate location time-series can be used by an adversary to build accurate profiles of users' movements. Finally, Boukoros et al.~\cite{boukoros2019johnny} study the effect of defenses on finding points of interest while computing aggregate statistics of geo-located measurements; in this work, we focus on a different privacy violation, i.e., membership inference.

\descr{Membership Inference on Aggregate Locations.} As discussed in Section~\ref{sec:priors},  Pyrgelis et al.~\cite{pyrgelis2018knock} %
model the problem of MIAs against aggregate locations using a distinguishability game, and train a classifier to differentiate aggregates including the data of a target from those that do not.
While our analysis is based on their attacks, our research objective is substantially different.
Pyrgelis et al.'s main goal is to investigate the feasibility of inference attacks;
whereas, we aim to gain a deeper understanding about the reasons behind the attacks' success, providing insights about locations and times that ease inference and the characteristics of the users that are affected more than others. %
Moreover,~\cite{pyrgelis2018knock} only studies the utility-privacy trade-off provided by differential privacy~\cite{rastogi2010differentially,dwork2008differential}, while we use the insights obtained in our analysis to select potential mitigation approaches, which we evaluate, both in terms of privacy and utility, in the context of various spatio-temporal analytics tasks. %

\descr{MIAs in Other Settings.} %
In~\cite{homer2008resolving}, Homer et al.~show that aggregate genomic statistics leak information about the inclusion of a target's genome in the dataset.
Then, Wang et al.~\cite{wang2009learning} improve on that by taking into account correlations within the human genome, while Backes et al.~\cite{backes2016membership} generalize the attack to other types of data like microRNA expression datasets. The data targeted by these studies does not have spatio-temporal dimensions, and thus they are orthogonal to our work.
Buscher et al.~\cite{bucher2017} study MIAs in the context of smart-metering, showing how aggregating a small number of household readings does not  protect the privacy of individual (house) profiles. While smart metering data is also a time-series, it does not have spatial components, and the correlations are different.

Finally, previous work studies membership inference on machine learning models, i.e., learning whether a data point was used to train a model, using the intuition that the model ends up overfitting on data used for training~\cite{shokri2017membership,yeom2018privacy}. The attack is also effective under less restrictive adversarial assumptions~\cite{DBLP:journals/corr/abs-1806-01246} and feasible in broader scenarios. For instance, Hayes et al.~\cite{hayes2017logan} show that MIAs are also possible against generative models, while Melis et al.~\cite{melis2018inference} do so for collaborative and federated learning. Jayaraman and Evans~\cite{jayaraman2019evaluating} study the impact of different variants of differential privacy, variable choices of the $\epsilon$ parameter, and several learning tasks on both utility and privacy (including in the context of MIAs) for privacy-preserving machine learning.
Overall, protection mechanisms against MIA on machine learning, such as dropout or model stacking~\cite{DBLP:journals/corr/abs-1806-01246}, or adversarial training~\cite{nasr2018machine}, cannot be applied in our scenario since the publication of aggregate location statistics does not involve learning. Thus, studies that aim at understanding why membership inference against machine learning works~\cite{demyst2018,long2018understanding} cannot inform defenses for aggregate location time-series.

\section{Conclusion}\label{sec:conclusion}

In this paper, we conducted an in-depth measurement study of why and in what settings membership inference attacks on aggregate location time-series are successful.
We found that both regular and uncommon mobility patterns are the easiest to recognize. 
Also, size matters: users contributing a lot of data to the aggregates are easier to attack. 
However, there is no characteristic that can be singled out and eliminated to thwart the attack on all fronts.

Then, our extensive measurements of the performance of defense strategies with respect to a range of applications confirmed that there does not exist a unique generic defense that protects against MIA and preserves the utility of the analytics for arbitrary applications. Nonetheless, we identified some strategies that provide reasonable utility for specific tasks with good average privacy protection. For instance, suppressing small counts can be used for ranking hotspots, data generalization for forecasting traffic, hotspot discovery, and map inference, while sampling is effective for location labeling and anomaly detection when the dataset is sparse. Our experiments also revealed that differentially private mechanisms only ``work'' (i.e., maintain utility at reasonably high levels of privacy) for some applications and more so when using weaker privacy notions like crowd-blending privacy. This is somewhat different from other settings where DP techniques can be fine-tuned for the safe release of aggregate statistics (e.g., the US Census data~\cite{abowd2018us} or telemetry data~\cite{erlingsson2014rappor}), as the very nature of location data as well as the high dimensionality of location time-series make it hard to protect everyone.

Overall, our measurement study, while comprehensive, is inevitably limited to the employed datasets, the considered attacks and defenses as well as their parameters. Thus, it remains an open question whether our results generalize to other location traces or different settings. To this end, as part of future work, we plan to investigate analytical methods that can be employed to study privacy--utility tradeoffs on aggregate location time-series and extend our experiments to other datasets, consider additional attacks on location aggregates other than MIA, and propose novel defense strategies that can be employed to thwart them while preserving the utility of the statistics for mobility analytics. In particular, a promising option is the use of differentially private noise highly tuned for specific tasks, while an additional avenue to explore is the generation of synthetic data tailored to mobility analytics that rely on aggregates instead of trajectories, i.e., schemes that preserve space and time and granularity while still providing privacy.

\bibliographystyle{ACM-Reference-Format} %
\bibliography{bibfile}

\appendix

\section{Utility Metrics for Defenses}\label{app:utility-tables}

We report tables that demonstrate how each defense affects the various utility metrics that we consider. For ease of comparisons, Table~\ref{table:tfl-sfc-random-guess} shows, for each of the considered metrics, the utility corresponding to a random guess.
\begin{table}[H]
\small
\begin{center}
\setlength{\tabcolsep}{3pt}
\begin{tabular}{ c | c c c c c c }
\textbf{Dataset} & \textbf{MRE} & \textbf{MRE 10\%} & \textbf{F1} & \textbf{$\tau$} & \textbf{JS} & \textbf{$r$} \\ \midrule
\textbf{TFL}  & 4285.809 &13.267 & 0.099  & 0.002  & 0.733  & -0.002 \\ 
\textbf{SFC} & 85.122 & 0.08 & 0.094 & -0.001 & 0.472  & 0.001 \\ 
\end{tabular}
\end{center}
\caption{Utility metrics corresponding to a random guess.}
\label{table:tfl-sfc-random-guess}
\vspace{-0.6cm}
\end{table}
\begin{table}[H]
\small
\begin{center}
\setlength{\tabcolsep}{3pt}
\begin{tabular}{ c | c c c c c c }
\textbf{Group Size} & \textbf{MRE} & \textbf{MRE 10\%} & \textbf{F1} & \textbf{$\tau$} & \textbf{JS} & \textbf{$r$} \\ \midrule
\textbf{5}   & 26.326   & 0.113 & 0.485 & 0.098 & 0.596 & 0.581 \\
\textbf{10} & 61.457   & 0.234 & 0.383 & 0.102 & 0.662 & 0.525 \\
\textbf{20} & 145.056 & 0.449 & 0.259 & 0.043 & 0.702 & 0.489 \\
\end{tabular}
\end{center}
\caption{TFL, Utility for Spatial Generalization (SPG).}
\label{table:tfl-spatial}
\vspace{-0.6cm}
\end{table}

\begin{table}[H]
\begin{center}
\small
\setlength{\tabcolsep}{3pt}
\begin{tabular}{ c | c c c c c c }
\textbf{Grid Size} & \textbf{MRE} & \textbf{MRE 10\%} & \textbf{F1} & \textbf{$\tau$} & \textbf{JS} & \textbf{$r$} \\ \midrule
\textbf{5x5} & 0.840     & 0.019 & 0.534 & 0.036 & 0.215 & 0.684 \\
\textbf{2x2} & 71.270   & 0.055 & 0.344 & 0.016 & 0.402 & 0.507 \\
\textbf{1x1} & 114.199 & 0.086 & 0.049 & 0.017 & 0.434 & 0.367 \\
\end{tabular}
\end{center}
\caption{SFC, Utility for Spatial Generalization (SPG).}
\label{table:sfc-spatial}
\vspace{-0.6cm}
\end{table}

\begin{table}[H]
\small
\begin{center}
\setlength{\tabcolsep}{3pt}
\begin{tabular}{ c | c c c c c c }
\textbf{Timeslot} & \textbf{MRE} & \textbf{MRE 10\%} & \textbf{F1} & \textbf{$\tau$} & \textbf{JS} & \textbf{$r$} \\ \midrule
\textbf{4 Hours}      & 0.146   & 0.152  & 0.776 & 0.076  & 0.272  & 0.596 \\ 
\textbf{8 Hours}      & 0.308   & 0.308  & 0.753 & 0.049  & 0.426  & 0.474 \\ 
\textbf{1 Day}         & 0.777   & 0.741  & 0.738 & 0.020  & 0.605  & 0.225 \\ 
\textbf{1 Week}      & 2.945   & 2.221  & 0.651 & 0.003  & 0.658  & 0.000 \\ 
\end{tabular}
\end{center}
\caption{TFL, Utility for Temporal Generalization (TG).}
\label{table:tfl-temporal}
\vspace{-0.6cm}
\end{table}

\begin{table}[H]
\small
\begin{center}
\setlength{\tabcolsep}{3pt}
\begin{tabular}{ c | c c c c c c }
\textbf{Timeslot} & \textbf{MRE} & \textbf{MRE 10\%} & \textbf{F1} & \textbf{$\tau$} & \textbf{JS} & \textbf{$r$} \\ \midrule
\textbf{4 Hours} & 0.134   & 0.066 & 0.823 & 0.056  &0.139  & 0.632 \\ 
\textbf{8 Hours} & 0.280   & 0.104 & 0.781 & 0.074  & 0.186 & 0.413 \\
\textbf{1 Day}    & 0.703   & 0.135 & 0.709 & 0.040  & 0.246 & 0.086 \\
\textbf{1 Week} & 2.493   & 0.149 & 0.455 & -0.057 & 0.324 & 0.000 \\ 
\end{tabular}
\end{center}
\caption{SFC, Utility for Temporal Generalization (TG).}
\label{table:sfc-temporal}
\vspace{-0.6cm}
\end{table}

\begin{table}[H]
\small
\begin{center}
\setlength{\tabcolsep}{3pt}
\begin{tabular}{ c | c c c c c c }
\textbf{$x$} & \textbf{MRE} & \textbf{MRE 10\%} & \textbf{F1} & \textbf{$\tau$} & \textbf{JS} & \textbf{$r$} \\ \midrule
\textbf{2}       & 0.884       & 0.002   & 0.955 & 0.353 & 0.108 & 0.640 \\
\textbf{5}       & 1.779       & 0.004   & 0.916 & 0.224 & 0.153 & 0.393 \\
\textbf{10}     & 4.472       & 0.008   & 0.853 & 0.182 & 0.238 & 0.278 \\
\textbf{50}     & 22.516     & 0.037   & 0.431 & 0.150 & 0.453 & 0.040 \\
\textbf{150}   & 67.673     & 0.153   & 0.399 & 0.151 & 0.578 & 0.003 \\
\textbf{9500} & 4281.495 & 13.261 & 0.400 & 0.151 & 0.724 & 0.000 \\
\end{tabular}
\end{center}
\caption{TFL, Utility for Data Generalization with Fixed Ranges of size $x$ (DGFR).}
\label{table:tfl-ranges}
\vspace{-0.6cm}
\end{table}

\begin{table}[H]
\small
\begin{center}
\setlength{\tabcolsep}{3pt}
\begin{tabular}{ c | c c c c c c }
\textbf{$x$} & \textbf{MRE} & \textbf{MRE 10\%} & \textbf{F1} & \textbf{$\tau$} & \textbf{JS} & \textbf{$r$} \\ \midrule
\textbf{2}     & 0.659    &  0.000 & 0.983  & 0.802 & 0.070 & 0.809 \\
\textbf{5}     & 1.358    &  0.001 & 0.953  & 0.525 & 0.103 & 0.700 \\
\textbf{10}   & 3.385    &  0.002 & 0.911  & 0.316 & 0.164 & 0.594 \\
\textbf{50}   & 17.012  &  0.011 & 0.658  & 0.084 & 0.354 & 0.239 \\ 
\textbf{100} & 34.214  &  0.020 & 0.218  & 0.031 & 0.416 & 0.059 \\
\textbf{250} & 84.667  &  0.054 & 0.103  & 0.017 & 0.432 & 0.000 \\
\end{tabular}
\end{center}
\caption{SFC, Utility for Data Generalization with Fixed Ranges of size $x$ (DGFR).}
\label{table:sfc-ranges}
\vspace{-0.6cm}
\end{table}

\begin{table}[H]
\small
\begin{center}
\setlength{\tabcolsep}{3pt}
\begin{tabular}{ c | c c c c c c }
\textbf{$x'$} & \textbf{MRE} & \textbf{MRE 10\%} & \textbf{F1} & \textbf{$\tau$} & \textbf{JS} & \textbf{$r$} \\ \midrule
\textbf{1}  & 0.136 & 0.041 & 0.711 & 0.043 & 0.175 & 0.439 \\ 
\textbf{2}  & 0.022 & 0.025 & 0.715 & 0.045 & 0.117 & 0.810 \\
\textbf{4}  & 0.005 & 0.012 & 0.785 & 0.060 & 0.068 & 0.889 \\
\textbf{8}  & 0.002 & 0.006 & 0.820 & 0.095 & 0.039 & 0.901 \\
\textbf{16}& 0.000 & 0.002 & 0.850 & 0.159 & 0.020 & 0.903 \\
\end{tabular}
\end{center}
\caption{TFL, Utility for Data Generalization with $x'$ Adaptive Ranges (DGAR).}
\label{table:tfl-ranges-loc}
\vspace{-0.6cm}
\end{table}

\begin{table}[H]
\small
\begin{center}
\setlength{\tabcolsep}{3pt}
\begin{tabular}{ c | c c c c c c }
\textbf{$x'$} & \textbf{MRE} & \textbf{MRE 10\%} & \textbf{F1} & \textbf{$\tau$} & \textbf{JS} & \textbf{$r$} \\ \midrule
\textbf{1}   & 0.045 & 0.021  & 0.745 & 0.083 & 0.153 & 0.412 \\ 
\textbf{2}   & 0.015 & 0.013  & 0.783 & 0.106 & 0.113 & 0.842 \\ 
\textbf{4}   & 0.006 & 0.006  & 0.873 & 0.167 & 0.063 & 0.943 \\
\textbf{8}   & 0.002 & 0.003  & 0.928 & 0.327 & 0.030 & 0.971 \\
\textbf{16} & 0.001 & 0.001  & 0.961 & 0.529 & 0.014 & 0.978 \\
\end{tabular}
\end{center}
\caption{SFC, Utility for Data Generalization with $x'$ Adaptive Ranges (DGAR).}
\label{table:sfc-ranges-loc}
\vspace{-0.6cm}
\end{table}

\begin{table}[H]
\small
\begin{center}
\setlength{\tabcolsep}{3pt}
\begin{tabular}{ c | c c c c c c }
\textbf{$k$} & \textbf{MRE} & \textbf{MRE 10\%} & \textbf{F1} & \textbf{$\tau$} & \textbf{JS} & \textbf{$r$} \\ \midrule
\textbf{2}   & 0.016 & 0.000 & 0.978 & 0.905 & 0.039 & 0.646 \\ 
\textbf{5}   & 0.028 & 0.000 & 0.962 & 0.818 & 0.078 & 0.399 \\
\textbf{10} & 0.035 & 0.002 & 0.915 & 0.618 & 0.119 & 0.285 \\ 
\textbf{20} & 0.041 & 0.010 & 0.767 & 0.402 & 0.171 & 0.162 \\ 
\textbf{40} & 0.045 & 0.038 & 0.485 & 0.259 & 0.224 & 0.061 \\
\textbf{80} & 0.047 & 0.056 & 0.398 & 0.161 & 0.256 & 0.012 \\
\end{tabular}
\end{center}
\caption{TFL, Utility for Suppressing Small Counts (SSC).}
\label{table:tfl-sup-counts}
\vspace{-0.6cm}
\end{table}

\begin{table}[H]
\small
\begin{center}
\setlength{\tabcolsep}{3pt}
\begin{tabular}{ c | c c c c c c }
\textbf{$k$} & \textbf{MRE} & \textbf{MRE 10\%} & \textbf{F1} & \textbf{$\tau$} & \textbf{JS} & \textbf{$r$} \\ \midrule
\textbf{2}     & 0.007  & 0.000 & 1.000 &  0.998 & 0.034 & 0.815 \\ 
\textbf{5}     & 0.015  & 0.000 & 0.999 &  0.992 & 0.087 & 0.710 \\
\textbf{10}   & 0.021  & 0.000 & 0.995 &  0.964 & 0.150 & 0.610 \\ 
\textbf{20}   & 0.029  & 0.001 & 0.978 &  0.885 & 0.269 & 0.503 \\ 
\textbf{40}   & 0.041  & 0.005 & 0.869 &  0.769 & 0.468 & 0.324 \\ 
\textbf{80}   & 0.050  & 0.032 & 0.381 &  0.214 & 0.532 & 0.104 \\ 
\textbf{160} & 0.053  & 0.059 & 0.103 &  0.015 & 0.595 & 0.005 \\ 
\end{tabular}
\end{center}
\caption{SFC, Utility for Suppressing Small Counts (SSC).}
\label{table:sfc-sup-counts}
\vspace{-0.6cm}
\end{table}

\begin{table}[H]
\small
\begin{center}
\setlength{\tabcolsep}{3pt}
\begin{tabular}{ c | c c c c c c }
\textbf{$z$} & \textbf{MRE} & \textbf{MRE 10\%} & \textbf{F1} & \textbf{$\tau$} & \textbf{JS} & \textbf{$r$} \\ \midrule
\textbf{0.2} & 0.000 & 0.0 & 1.000 & 1.000 & 0.000 & 0.904 \\ 
\textbf{0.4} & 0.001 & 0.0 & 0.999 & 0.973 & 0.002 & 0.892 \\ 
\textbf{0.6} & 0.006 & 0.0 & 0.991 & 0.896 & 0.018 & 0.818 \\ 
\textbf{0.8} & 0.021 & 0.0 & 0.968 & 0.576 & 0.076 & 0.639 \\ 
\end{tabular}
\end{center}
\caption{TFL, Utility for Suppressing Less Popular Locations/Timeslots (SLP).}
\label{table:tfl-sup-locs-ts}
\vspace{-0.6cm}
\end{table}

\begin{table}[H]
\small
\begin{center}
\setlength{\tabcolsep}{3pt}
\begin{tabular}{ c | c c c c c c }
\textbf{$z$} & \textbf{MRE} & \textbf{MRE 10\%} & \textbf{F1} & \textbf{$\tau$} & \textbf{JS} & \textbf{$r$} \\ \midrule
\textbf{0.2} & 0.001 & 0.0 & 1.000 & 1.000 & 0.002 & 0.968 \\ 
\textbf{0.4} & 0.006 & 0.0 & 0.999 & 0.987 & 0.039 & 0.891 \\ 
\textbf{0.6} & 0.016 & 0.0 & 0.989 & 0.905 & 0.149 & 0.717 \\ 
\textbf{0.8} & 0.031 & 0.0 & 0.951 & 0.797 & 0.336 & 0.523 \\ 
\end{tabular}
\end{center}
\caption{SFC, Utility for Suppressing Less Popular Locations/Timeslots (SLP).}
\label{table:sfc-sup-locs-ts}
\vspace{-0.6cm}
\end{table}

\begin{table}[H]
\small
\begin{center}
\setlength{\tabcolsep}{3pt}
\begin{tabular}{ c | c c c c c c }
\textbf{$w$} & \textbf{MRE} & \textbf{MRE 10\%} & \textbf{F1} & \textbf{$\tau$} & \textbf{JS} & \textbf{$r$} \\ \midrule
\textbf{0.2} & 0.010 & 0.012 & 0.939 & 0.222 & 0.029 & 0.838 \\ 
\textbf{0.4} & 0.019 & 0.024 & 0.905 & 0.190 & 0.047 & 0.759 \\ 
\textbf{0.6} & 0.029 & 0.036 & 0.863 & 0.173 & 0.068 & 0.656 \\ 
\textbf{0.8} & 0.038 & 0.048 & 0.792 & 0.160 & 0.105 & 0.503 \\ 
\end{tabular}
\end{center}
\caption{TFL, Utility for Sampling (SMP).}
\label{table:tfl-sampling}
\vspace{-0.6cm}
\end{table}
\begin{table}[H]
\small
\begin{center}
\setlength{\tabcolsep}{3pt}
\begin{tabular}{ c | c c c c c c }
\textbf{$w$} & \textbf{MRE} & \textbf{MRE 10\%} & \textbf{F1} & \textbf{$\tau$} & \textbf{JS} & \textbf{$r$} \\ \midrule
\textbf{0.2} & 0.011 & 0.012 & 0.931 & 0.329 & 0.040 & 0.922  \\
\textbf{0.4} & 0.021 & 0.024 & 0.894 & 0.208 & 0.066 & 0.849  \\ 
\textbf{0.6} & 0.032 & 0.036 & 0.846 & 0.145 & 0.099 & 0.751 \\
\textbf{0.8} & 0.042 & 0.048 & 0.761 & 0.099 & 0.160 & 0.594  \\
\end{tabular}
\end{center}
\caption{SFC, Utility for Sampling (SMP).}
\label{table:sfc-sampling}
\vspace{-0.6cm}
\end{table}
\begin{table}[H]
\small
\begin{center}
\setlength{\tabcolsep}{3pt}
\begin{tabular}{ c | c c c c c c }
\textbf{$k$} & \textbf{MRE} & \textbf{MRE 10\%} & \textbf{F1} & \textbf{$\tau$} & \textbf{JS} & \textbf{$r$} \\ \midrule
\textbf{2}    & 0.434 & 0.000 & 0.829 & 0.646 & 0.080  & 0.570 \\ 
\textbf{5}    & 0.439 & 0.000 & 0.824 & 0.584 & 0.084  & 0.553 \\ 
\textbf{10}  & 0.440 & 0.001 & 0.815 & 0.449 & 0.085  & 0.548 \\ 
\textbf{20}  & 0.440 & 0.001 & 0.801 & 0.309 & 0.086  & 0.546 \\ 
\textbf{40}  & 0.439 & 0.002 & 0.795 & 0.201 & 0.086  & 0.546 \\ 
\textbf{80}  & 0.439 & 0.002 & 0.795 & 0.159 & 0.086  & 0.546 \\
\end{tabular}
\end{center}
\caption{TFL, Utility for Perturbing Small Counts with $\epsilon'{=}1.0$ (PSC).}
\label{table:tfl-cb2}
\vspace{-0.6cm}
\end{table}
\begin{table}[H]
\small
\begin{center}
\setlength{\tabcolsep}{3pt}
\begin{tabular}{ c | c c c c c c }
\textbf{$k$} & \textbf{MRE} & \textbf{MRE 10\%} & \textbf{F1} & \textbf{$\tau$} & \textbf{JS} & \textbf{$r$} \\ \midrule
\textbf{2}     & 3.397 & 0.000 & 0.961 & 0.810 & 0.163 & 0.610 \\
\textbf{5}     & 3.411 & 0.000 & 0.949 & 0.780 & 0.185 & 0.567 \\ 
\textbf{10}   & 3.427 & 0.000 & 0.930 & 0.749 & 0.204 & 0.534 \\ 
\textbf{20}   & 3.478 & 0.000 & 0.897 & 0.696 & 0.228 & 0.474 \\ 
\textbf{40}   & 3.448 & 0.002 & 0.840 & 0.546 & 0.249 & 0.417 \\ 
\textbf{80}   & 3.417 & 0.006 & 0.754 & 0.217 & 0.255 & 0.395 \\ 
\textbf{160} & 3.429 & 0.009 & 0.746 & 0.116 & 0.255 & 0.393 \\ 
\end{tabular}
\end{center}
\caption{SFC, Utility for Perturbing Small Counts with $\epsilon'{=}0.1$ (PSC).}
\label{table:sfc-cb2}
\vspace{-0.6cm}
\end{table}
\begin{table}[H]
\small
\begin{center}
\setlength{\tabcolsep}{3pt}
\begin{tabular}{ c | c c c c c c }
\textbf{$\epsilon$} & \textbf{MRE} & \textbf{MRE 10\%} & \textbf{F1} & \textbf{$\tau$} & \textbf{JS} & \textbf{$r$} \\ \midrule
\textbf{0.01} & 112.526 & 0.381 & 0.124 & 0.034 & 0.666 & 0.005 \\
\textbf{0.1}   & 11.310   & 0.058 & 0.251 & 0.032 & 0.412 & 0.043 \\
\textbf{1.0}   & 1.149     & 0.022 & 0.721 & 0.043 & 0.175 & 0.236 \\
\textbf{10.0} & 0.087     & 0.020 & 0.758 & 0.053 & 0.116 & 0.447 \\
\end{tabular}
\end{center}
\caption{TFL, Utility for Fourier Perturbation Algorithm (FPA).}
\label{table:tfl-fpa}
\vspace{-0.6cm}
\end{table}
\begin{table}[H]
\small
\begin{center}
\setlength{\tabcolsep}{3pt}
\begin{tabular}{ c | c c c c c c }
\textbf{$\epsilon$} & \textbf{MRE} & \textbf{MRE 10\%} & \textbf{F1} & \textbf{$\tau$} & \textbf{JS} & \textbf{$r$} \\ \midrule
\textbf{0.01}  & 74.957 & 0.106 & 0.136 & 0.047 & 0.577 & 0.001 \\ 
\textbf{0.1}    & 22.855 & 0.052 & 0.220 & 0.008 & 0.523 & 0.034 \\ 
\textbf{1.0}    & 2.387   & 0.013 & 0.784 & 0.093 & 0.218 & 0.263 \\
\textbf{10.0}  & 0.234   & 0.011 & 0.841 & 0.137 & 0.107 & 0.540 \\
\end{tabular}
\end{center}
\caption{SFC, Utility for Fourier Perturbation Algorithm (FPA).}
\label{table:sfc-fpa}
\vspace{-0.6cm}
\end{table}
\begin{table}[H]
\small
\begin{center}
\setlength{\tabcolsep}{3pt}
\begin{tabular}{ c | c c c c c c }
\textbf{$w$} & \textbf{MRE} & \textbf{MRE 10\%} & \textbf{F1} & \textbf{$\tau$} & \textbf{JS} & \textbf{$r$} \\ \midrule
\textbf{0.2} & 0.444 & 0.012 & 0.777 & 0.091 & 0.099 & 0.505 \\
\textbf{0.4} & 0.449 & 0.024 & 0.741 & 0.066 & 0.119 & 0.448 \\
\textbf{0.6} & 0.454 & 0.036 & 0.688 & 0.054 & 0.150 & 0.374 \\
\textbf{0.8} & 0.460 & 0.048 & 0.570 & 0.042 & 0.214 & 0.261 \\
\end{tabular}
\end{center}
\caption{TFL, Utility for Sampling \& Perturbing Small Counts with $k{=}5$ and $\epsilon'{=}1.0$ (SPSC).}
\label{table:tfl-zkp}
\vspace{-0.6cm}
\end{table}
\begin{table}[H]
\small
\begin{center}
\setlength{\tabcolsep}{3pt}
\begin{tabular}{ c | c c c c c c }
\textbf{$w$} & \textbf{MRE} & \textbf{MRE 10\%} & \textbf{F1} & \textbf{$\tau$} & \textbf{JS} & \textbf{$r$} \\ \midrule
\textbf{0.2} & 3.348 & 0.012 & 0.812 & 0.248  & 0.263 & 0.410 \\
\textbf{0.4} & 3.394 & 0.024 & 0.698 & 0.122  & 0.311 & 0.333 \\
\textbf{0.6} & 3.388 & 0.036 & 0.504 & 0.037  & 0.379 & 0.237 \\
\textbf{0.8} & 3.459 & 0.047 & 0.279 & -0.011 & 0.481 & 0.115 \\
\end{tabular}
\end{center}
\caption{SFC, Utility for Sampling \& Perturbing Small Counts with $k{=}20$ and $\epsilon'{=}0.1$ (SPSC).}
\label{table:sfc-zkp}
\vspace{-0.6cm}
\end{table}
\begin{table}[H]
\small
\begin{center}
\setlength{\tabcolsep}{3pt}
\begin{tabular}{ c | c c c c c c }
\textbf{$w$} & \textbf{MRE} & \textbf{MRE 10\%} & \textbf{F1} & \textbf{$\tau$} & \textbf{JS} & \textbf{$r$} \\ \midrule
\textbf{0.2} & 0.033 & 0.025 & 0.757 & 0.052 & 0.113 & 0.452 \\
\textbf{0.4} & 0.036 & 0.033 & 0.754 & 0.051 & 0.115 & 0.419 \\
\textbf{0.6} & 0.040 & 0.041 & 0.750 & 0.051 & 0.118 & 0.372 \\
\textbf{0.8} & 0.044 & 0.050 & 0.736 & 0.046 & 0.129 & 0.290 \\
\end{tabular}
\end{center}
\caption{TFL, Utility for Sampling \& Fourier Perturbation Algorithm with $\epsilon{=}10.0$ (SFPA).}
\label{table:tfl-sampling-fpa}
\vspace{-0.6cm}
\end{table}
\begin{table}[H]
\small
\begin{center}
\setlength{\tabcolsep}{3pt}
\begin{tabular}{ c | c c c c c c }
\textbf{$w$} & \textbf{MRE} & \textbf{MRE 10\%} & \textbf{F1} & \textbf{$\tau$} & \textbf{JS} & \textbf{$r$} \\ \midrule
\textbf{0.2} & 0.214 & 0.016 & 0.839 & 0.137 & 0.109 & 0.526 \\
\textbf{0.4} & 0.192 & 0.025 & 0.838 & 0.135 & 0.112 & 0.506 \\
\textbf{0.6} & 0.159 & 0.036 & 0.834 & 0.129 & 0.119 & 0.477 \\
\textbf{0.8} & 0.109 & 0.048 & 0.824 & 0.111 & 0.134 & 0.413 \\
\end{tabular}
\end{center}
\caption{SFC, Utility for Sampling \& Fourier Perturbation Algorithm with $\epsilon{=}10.0$ (SFPA).}
\label{table:sfc-sampling-fpa}
\vspace{-0.6cm}
\end{table}
\end{document}